\documentstyle[12pt,epsfig]{article}

\newcommand{\bara}{\begin{array}{c}}
\newcommand{\eara}{\end{array}}

\def\be{\begin{equation}}
\def\ee{\end{equation}}
\def\ba{\begin{eqnarray}}
\def\ea{\end{eqnarray}}
\def\bann{\begin{eqnarray*}}
\def\eann{\end{eqnarray*}}
\def\benn{\begin{displaymath}}
\def\eenn{\end{displaymath}}
\def\nn{\nonumber}
\def\lapproxeq{{\ \lower 0.6ex \hbox{$\buildrel<\over\sim$}\ }}
\def\gapproxeq{{\ \lower 0.6ex \hbox{$\buildrel>\over\sim$}\ }}
\begin{document}  
\vspace*{-2cm}  
\renewcommand{\thefootnote}{\fnsymbol{footnote}}  
\begin{flushright}  
hep-ph/9902234\\
DTP/99/02\\
February 1999\\  
\end{flushright}  
\vskip 65pt  
\begin{center}  
{\Large \bf Sudakov Logarithm Resummation in Transverse
Momentum Space for Electroweak Boson Production at Hadron Colliders}\\
\vspace{1.2cm} 
{\bf  
Anna~Kulesza${}^1$\footnote{Anna.Kulesza@durham.ac.uk} and  
W.~James~Stirling${}^{1,2}$\footnote{W.J.Stirling@durham.ac.uk}  
}\\  
\vspace{10pt}  
{\sf 1) Department of Physics, University of Durham,  
Durham DH1 3LE, U.K.\\  
  
2) Department of Mathematical Sciences, University of Durham,  
Durham DH1 3LE, U.K.}  
  
\vspace{70pt}  
\begin{abstract}
A complete description of $W$ and $Z$ boson
production at high-energy hadron colliders requires the resummation
of large Sudakov double logarithms which dominate the transverse momentum
($q_T$) distribution at small $q_T$. We compare different prescriptions
for performing this resummation, in particular implicit impact parameter
space resummation versus explicit transverse momentum space resummation.
We argue that the latter method can be formulated so as to retain the 
advantages of the former, while at the same time allowing a smooth
transition to finite order dominance at high $q_T$.
\end{abstract}
\end{center}  
\vskip12pt

\setcounter{footnote}{0}  
\renewcommand{\thefootnote}{\arabic{footnote}}  
\vfill  
\clearpage  
\setcounter{page}{1}  
\pagestyle{plain} 
\section{Introduction}
\label{sec:intro}

The production of $W$  and $Z$ bosons  at 
hadron collliders provides several fundamental tests
 of perturbative QCD. 
A problem of particular topical interest is the calculation of
 the transverse momentum ($q_T$)
distribution of the produced vector boson. Data from the Tevatron
$p\bar p$ collider 
experiments  now cover the  regions of both small and large $q_T$ 
\cite{datareview}. For large
$q_T$, fixed-order perturbative  calculations  should be sufficient, and indeed
the current ${\cal{O}}(\alpha^2_S)$ predictions  agree well  with experiment. 
However, it is the small $q_T$ region which is more
theoretically challenging as here
one encounters the infrared structure of the theory in the form of large
logarithms of $Q/q_T$, where $Q$ is of the order  of the weak boson
mass~\cite{DDT}. The presence of higher-order contributions of order
$\alpha_S^n \ln^{2n-1}(Q^2/q_T^2)$ leads to a breakdown of fixed-order
perturbation theory as $q_T \to 0$.
Although
there is much data available, a completely consistent theoretical
treatment in agreement with experiment has not yet been developed.\footnote{For 
a review of the literature, see for example \cite{Ellis:Ross:Veseli}.}
Furthermore the most sophisticated treatments~\cite{b-space}, involving the
resummation of the large logarithms in impact parameter space, lead to results
that are in practice difficult to merge with the large $q_T$ fixed-order 
expressions.
The form of the low $q_T$ distribution is not only of theoretical
interest -- for example, a proper description is needed for an accurate determination 
of the $W$ boson mass. The formalism also applies directly
 to the production of {\it any} massive colour-neutral
particle, including the Higgs boson.

The purpose of this paper is to investigate the resummation of large
$\ln(\frac{Q}{q_T^2})$ logarithms directly in momentum space. The question
has been addressed recently in Refs.~\cite{Ellis:Ross:Veseli} and \cite{Ellis:Veseli}, 
and our 
analysis can be considered
an extension of these studies. 

We begin with a simple formulation of the problem.
The large logarithms in the $W$ transverse momentum distribution
\footnote{For simplicity we will only
refer to the $W$ boson in our study, but obviously 
all our conclusions apply equally
well to $Z$ production also.} at small $q_T$
arise from soft gluon emission from the incoming (massless) quarks in
the basic $q \bar{q}' \to W $ process.
The logarithmic dependence enters the differential cross section formula
 through terms
proportional to (for ${d \sigma \over d q_T^2}$)
\be
\alpha_S^n  \ln^{2n-m} \left( \frac{Q^2}{q_T^2} \right) \qquad
m=1,...,2n \ .
\label{logs}
\ee
In practice, the coefficients of the terms represented by the
expressions in~(\ref{logs}) are known only for $m=1,2,3,4$; the
rest remain unknown.
In the simplest case,  the cross section at small $q_T$ 
is approximated by summing
only the leading logarithm terms in (\ref{logs}) (i.e. $m=1$), which gives
 a Sudakov form factor (for the $q \bar{q}' \to W $ subprocess cross section): 
\be
{1 \over \sigma_0} {d \sigma \over d q_T^2}_{(DLLA)} = 
{ \alpha_S C_F \over \pi q_T^2} \ln \left({Q^2 \over q_T^2}\right)
\exp \left({- \alpha_S C_F \over 2 \pi}\ln^2 \left({Q^2 \over
q_T^2}\right)\right) \ .
\label{DLLA}
\ee
This double leading logarithm approximation (DLLA) corresponds to the situation when only
the contributions of soft and collinear gluons are included and the strong
ordering of the gluons' momenta is additionally imposed~\cite{Ellis:Stirling}. 
The fact that the Sudakov form factor in~(\ref{DLLA}) results in a {\it
  suppression} of the  cross section as $q_T \to 0$ indicates that sub-leading
logarithms (i.e. $m\geq 2$ 
in~(\ref{logs})) have to be taken into account in the resummation. Two methods of 
doing this have been proposed.
\begin{itemize}
\item[(i)]
Resummation in impact parameter $b$ space (where
$\vec{b}$ is a two-dimensional Fourier conjugate of $\vec{q}_T$). This
method correctly takes into account the known logarithmic terms and also certain
kinematic features of the gluon emission. However it suffers from several
drawbacks, see~\cite{Ellis:Ross:Veseli,Ellis:Veseli}, in particular it leads to an unphysical
extrapolation to large $q_T$, as  will be illustrated below.
\item[(ii)]
Resummation in $q_T$-space (proposed recently in Ref.~\cite{Ellis:Veseli}).
This method includes exactly the same logarithms as the  $b$-space formalism 
up to and including 
 $\alpha_S^n \ln^{2n-3}(\frac{Q^2}{q_T^2})$ terms, but omits sub-leading
 `kinematic' logarithms, i.e. sub-leading logarithms which arise
 when the strong ordering assumption is relaxed 
 and which are automatically included in
 $b$-space. These terms start to contribute from the fourth `tower' of
 logarithms down onwards. The question is whether it is possible to include sufficient kinematic logarithms using
this technique that the $b$-space cross section can be adequately
approximated by resummation in $q_T$ space {\it in the region of
$q_T$ relevant to the comparison with data}. Furthermore, in this approach
the problems with merging the resummed and fixed-order results
in the `medium' $q_T$ region can be more easily circumvented.
\end{itemize}
In this paper we explore further  the $q_T$-space resummation approach.
We extend the work of \cite{Ellis:Veseli} by including all NNNL
logarithms and higher `towers' of known kinematic logarithms. The effect on
the cross section of systematically adding sub-leading terms is quantified.
The goal is to achieve a momentum-space-resummed cross section which 
reproduces the impact-parameter-resummed cross section in the regions
of $q_T$ relevant to the experimental measurements and which includes all
known calculated coefficients.

The paper is organized as follows.
In the following section we present the basic theoretical expressions
for the resummed cross section in both $b$ and $q_T$ spaces.
In Section~\ref{sec:quant} we consider in turn the quantitative effect
of the various sub-leading logarithm contributions, in particular
those from higher-order coefficients, kinematics, and the running 
of $\alpha_S$. 
Section~\ref{sec:conc} contains a summary and some conclusions.

\section{Theoretical description of the small $q_T$ distribution:
resummation in $q_T$ space}
\label{sec:theory}

The general $b$-space expression for the differential cross section for vector
boson production at small $q_T$ has been given in~\cite{Ellis:Ross:Veseli}
(and see references therein). 
For  simplicity, throughout this study 
we shall restrict our attention to the parton-level 
subprocess cross section: parton distribution functions can in principle
be incorporated to yield the hadron level cross section without changing any of our conclusions on resummation.\footnote{Formally, one can imagine taking
$(\tau = Q^2/s)^N$  {\it moments} of the hadron cross section, which allows the 
subprocess cross section to be factored out. Our subprocess 
cross section corresponds to the $N=0$ moment in this sense, $\frac{d
  \sigma}{d q_T^2}\equiv {1 \over q_T^2}\Sigma(0,{Q^2 \over q_T^2},{Q^2\over
  \mu^2},\alpha_S(\mu^2))$, see Ref.~\cite{Davies:Stirling}.}
Using the $b$-space formalism, at the purely perturbative level,
 one obtains for the differential cross section~\cite{Ellis:Ross:Veseli} 
\be
{d\sigma \over d q_T^2} ={\sigma_0 \over 2}\int_{0}^{\infty} b d b  \, J_{0}( q_{\tiny T}
b) e^{S(b,Q^2)} \,,
\label{resum}
\ee
where
\ba
\label{eq:abseries}
S(b,Q^2) = - \int_{b_0^2 \over b^2}^{Q^2} \frac{d\bar\mu^2}{\bar\mu^2} 
\bigg[ \ln \bigg ( {Q^2\over\bar \mu^2} \bigg ) A(\alpha_S(\bar\mu^2)) +
B(\alpha_S(\bar\mu^2)) \bigg ] \,,\label{Sbs} \\
A(\alpha_S) = \sum^\infty_{i=1} \left(\frac{\alpha_S}{2 \pi} \right)^i A^{(i)}\
, \quad
B(\alpha_S) = \sum^\infty_{i=1} \left(\frac{\alpha_S}{2 \pi} \right)^i B^{(i)}\
,\nn
\ea
\bann
b_0=2\exp(-\gamma_E)\,, \quad \sigma_0={4 \pi \alpha^2 \over 9 s} \,. 
\eann
and $Q=M_W$ in the present context.

\begin{figure}
\begin{center}
\mbox{\epsfig{figure=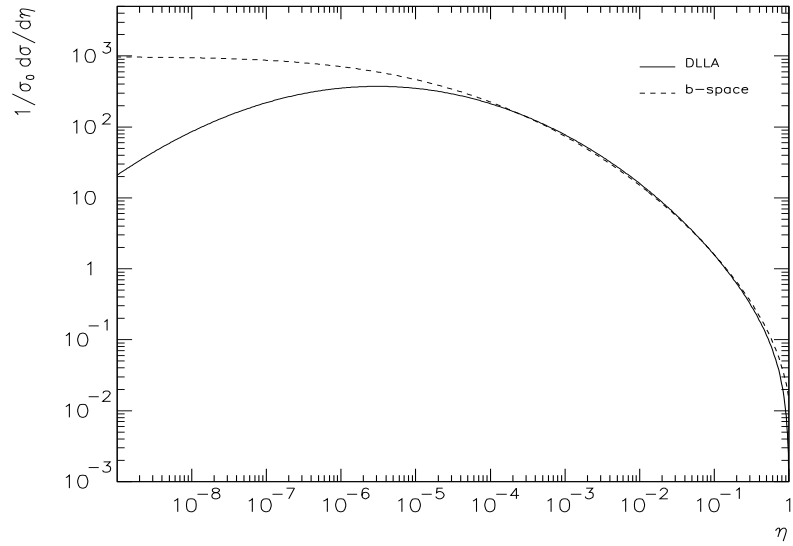,height=7.5cm,width=14cm}}
\end{center}
\caption{DLLA~(\ref{DLLA}) and $b$-space~(\ref{b_space}) results for the
  transverse momentum distribution ${1 \over \sigma_0} {d \sigma \over d \eta}$.}
\label{bspace_v_DLLA}
\end{figure}

\begin{figure}
\begin{center}
\mbox{\epsfig{figure=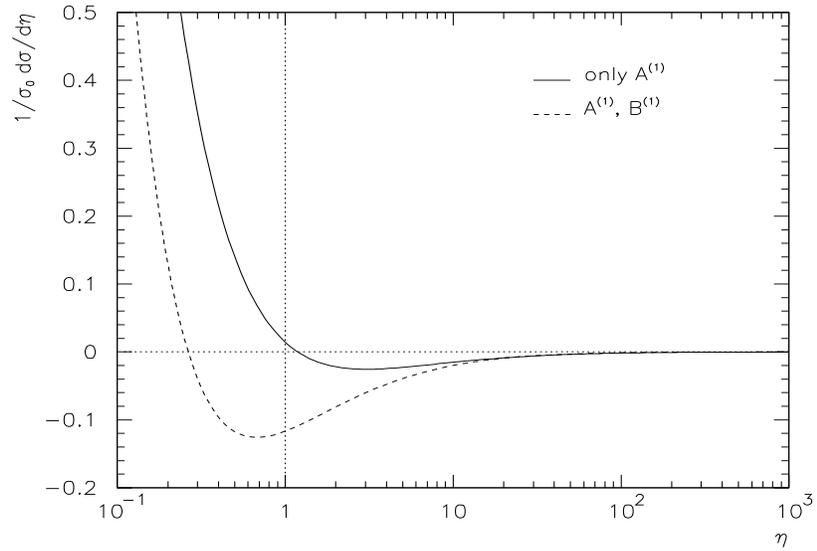,height=7.5cm,width=14cm}}
\end{center}
\caption{Extension of the $b$-space cross section presented in
  Fig.~\ref{bspace_v_DLLA} to large values of $\eta$.}
\label{wiggles}
\end{figure}

The first two coefficients in each of the series in~(\ref{eq:abseries}), i.e.
$A^{(1)},\ A^{(2)},\ B^{(1)},\ B^{(2)}$, are well
known~\cite{Davies:Stirling}. They can be obtained from the exact
LO $+$ NLO perturbative calculations in the
high $q_T$ region by comparing the logarithmic terms therein with 
the corresponding logarithms generated by  the first three terms of the expansion of
 $\exp(S(b,Q^2))$ in~(\ref{resum}). Explicitly,
\ba
A^{(1)}&=&2 C_F\ ,\nn\\
A^{(2)}&=&2 C_F
\Big( N (\frac{67}{18}-\frac{\pi^2}{6})-\frac{10}{9} T_R n_f \Big)
\ ,\nn \\
B^{(1)}&=&-3 C_F\ ,\nn \\
B^{(2)}&=& C_F^2 \Big(\pi^2-\frac{3}{4}-12 \zeta(3)\Big)
      +C_F N \Big( \frac{11}{9} \pi^2-\frac{193}{12}+6 \zeta(3)\Big)
\nn \\
&+&C_F T_R n_f \Big(\frac{17}{3}-\frac{4}{9} \pi^2\Big)  \ ,
\ea
 with $C_F=4/3$, $T_R=1/2$ and $N=3$.
It is instructive to see how the logarithms in $b$-space
generate logarithms in $q_T$-space. For illustration, we
take only the leading coefficient
$A^{(1)}=2C_F$ to be non-zero  in $e^{S(b,Q^2)}$, and assume a fixed coupling
$\alpha_S$. This corresponds to  
\be
{d\sigma \over d q_T^2} = {\sigma_0 \over 2}\int_{0}^{\infty} b d b \,  J_{0}( q_{\tiny T}  b) \exp
\bigg( -\frac{\alpha_s C_F}{2 \pi} \ln^2 \bigg ( \frac{Q^2 b^2}{b_0^2}
\bigg) \bigg) \ .
\ee 
The expressions are made more compact by defining
new variables $\eta={q_T^2 \over Q^2}$, $z=b^2 Q^2 $,
$\lambda={\alpha_S C_F \over \pi}$, \mbox{$z_0=4\exp(-2\gamma_E)=b_0^2$.} 
Then 
\ba
{1 \over \sigma_0} {d \sigma \over d \eta} = {1 \over 4} \int_{0}^{\infty} d z 
J_{0}( \sqrt{z\eta}) e^{-{\lambda \over 2} \ln^2
\big({z \over z_0} \big)}
\label{b_space}
\ea
and we encounter the same expression as in~\cite{Ellis:Fleishon:Stirling},
which describes the emission of soft and collinear gluons with
transverse momentum conservation taken into account.
The result of numerically integrating~(\ref{b_space}) and its comparison
with the DLLA approximation~(\ref{DLLA}) is shown 
in Fig.~\ref{bspace_v_DLLA}.\footnote{For all our studies with a 
fixed value of the coupling we take
    $\lambda=\frac{\alpha_S C_F}{\pi}=0.085$, $n_f=4$.}
The cross sections are similar over a broad range in $\eta$: the main differences
occur at (i) small $\eta$, where the DLLA curve is suppressed to zero and the 
$b$-space curve tends to a finite value, and (ii) at large $\eta$ (strictly,
outside the domain of validity of either expression). In the latter case,
the $q_T$-space cross section vanishes at $\eta=1$ by virtue of the overall
factor of $\ln(1/\eta)$. This is a crude approximation to the (formally correct)
vanishing of the leading-order cross section at the kinematic limit $\sim
M_W$.
In contrast, the $b$-space cross section has no information about this kinematic
limit, and is non-zero at $\eta = 1$. Furthermore, at large $q_T$ the 
$b$-space cross section {\it oscillates} about 0. This can be seen in Fig.~\ref{wiggles},
which extends the cross section of Fig.~\ref{bspace_v_DLLA} 
to large $\eta$ on a linear scale. The first zero of the oscillation is clearly evident.
Now since this occurs far outside the physical region it might be argued that
it is not a problem in practice. However, when the first sub-leading logarithm $B^{(1)}$
is included, the first zero moves {\it inside} the physical region, as shown in 
the figure. It is this behaviour which causes problems in merging the large $q_T$ fixed-order
result with the resummed expression, since the compensating terms have also
to be given an unphysical oscillating form.

After performing a partial integration and expanding terms under the 
integral in ~(\ref{b_space}), one obtains
\ba
{1 \over \sigma_0} {d \sigma \over d \eta} = 
{1 \over \eta} \sum_{N=1}^{\infty} \lambda^N { (-1/2)^{N-1} \over (N-1)!} 
\sum_{m=0}^{2N-1} { 2^m \bar{b}_m(\infty) } { \left( \begin{array}{c} 2N-1 \\ m \end{array} \right)}
\ln^{2N-1-m}\bigg ( {1 \over \eta} \bigg)
\label{qt_sum1}
\ea
which is now a perturbation series in $q_T$ space.
Here the numbers $\bar b_m(\infty)$ are defined by  
\be
\bar b_m(\infty) \equiv \int_{0}^{\infty} d y J_{1}(y) \ln^m({y \over
  \sqrt{z_0}})
\label{b_def}
\ee
and can be calculated explicitly from the generating function
\be
\sum_{m=0}^{\infty} {1 \over m!} t^m \bar{b}_m({\infty}) =
\exp \bigg[ -2 \sum_{k=1}^{\infty} { \zeta(2k+1) \over 2k+1} {\left( t \over
2 \right) }^{2k+1} \bigg] \,,
\label{b_form}
\ee
so that e.g. $\bar{b}_0({\infty})=1,\
\bar{b}_1({\infty})=\bar{b}_2({\infty})=0,\ \bar{b}_3({\infty})=-{1
\over 2}\zeta(3)$ etc.

Before studying the various approximations to the cross section in detail, it is worthwhile
comparing the two expressions for the differential cross section (\ref{b_space}) and (\ref{qt_sum1}). 
The first,
in $b$ space, calculates the cross section in terms of a one-dimensional integral.
The cross section is well  defined at all values of $q_T$, and in particular at $q_T = 0$.
In practice, however, non-perturbative (e.g. intrinsic $k_T$ smearing) effects dominate in
this region. These can be modelled by introducing an additional 
large-$b$ suppressing piece
in the integrand, for example $\exp(-\sigma^2 b^2)$ where $\sigma$ is a measure
of the $k_T$ smearing. In practice, since the argument 
of the running coupling in $b$-space is proportional to $1/b$, some form of
large-$b$ cut-off or freezing  must also be performed.

As already mentioned, however, the large $q_T$ behaviour of
(\ref{b_space}) is not physical: the integral has no knowledge of the exact kinematic upper limit
on $q_T$, although numerically it becomes small when $q_T \sim Q$. More problematically,
as $q_T$ is increased the distribution starts to {\it oscillate}, and it is this feature
(built-in via the Bessel function) which makes it difficult to merge with the finite-order
large $q_T$ cross section.

In contrast, the $q_T$-space cross section (\ref{qt_sum1}) is an {\it asymptotic} series.
The logarithms are singular at $q_T=0$, although as argued above this is in any case the region
where non-perturbative effects dominate. As with any asymptotic series, care must be taken
with the number of terms retained. Merging with the fixed-order large $q_T$ cross section is 
in principle straightforward,
since the $q_T$ logarithms  of (\ref{qt_sum1})  can simply be removed from the finite-order
pieces to avoid double counting. 

\section{Quantitative study of resummed cross sections}
\label{sec:quant}

The sub-leading corrections to~(\ref{DLLA}) have three origins. First, there
are sub-leading terms arising from the matrix elements that are associated with the
coefficients  $A^{(2)},B^{(1)}$, etc. Secondly, there are sub-leading effects
resulting from the running of the strong coupling $\alpha_S$. Finally, there
are also sub-leading terms in the form of kinematic logarithms, always
appearing with $\bar{b}_m(\infty)$ ($m\geq 1$) coefficients. In the following
subsection we focus on this particular type of sub-leading effect and assess
its importance. We isolate the effects induced by kinematic logarithms by
fixing the coupling and taking only leading terms arising from  the matrix
element. Then, in the following subsections, we subsequently switch on
running coupling effects and sub-leading logarithms from the matrix element.

\subsection{Fixed coupling analysis}

We begin our study of~(\ref{qt_sum1}) by performing the
 resummation for the simple case \mbox{$m=0$}. This is the DLLA of
 Eq.~(\ref{DLLA}), i.e. all radiated gluons are soft and
collinear with strong transverse momentum and energy ordering, and no account
taken of transverse momentum conservation:
\be
{1 \over \sigma_0} {d \sigma \over d \eta} = {1 \over \eta}
\sum_{N=1}^{\infty} \lambda^N { (-1/2)^{N-1} \over (N-1)!} \bar b_0(\infty)
\ln^{2N-1} \bigg({1\over \eta} \bigg) \,.
\label{sum_DLLA}
\ee

\begin{figure}[p]
\begin{center}
\mbox{\epsfig{figure=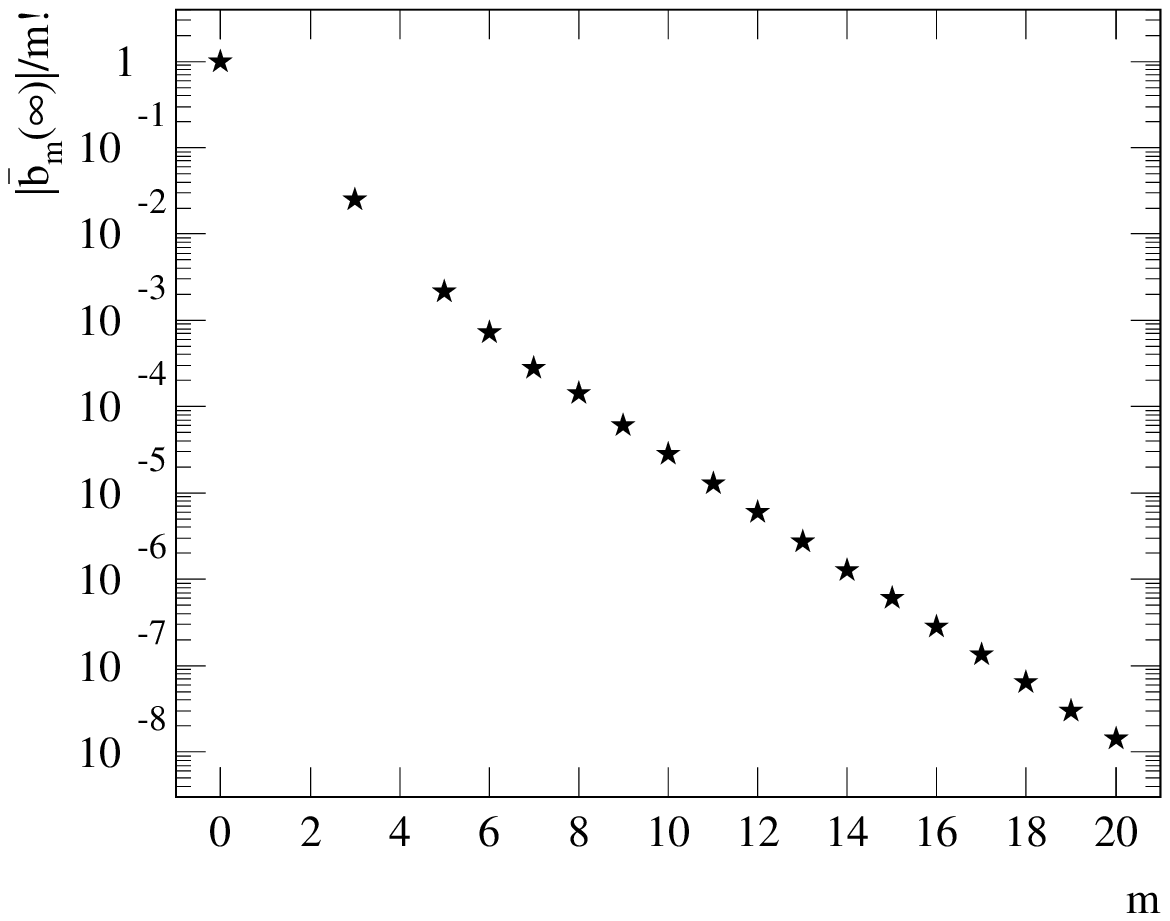,height=7.5cm,width=14cm}}
\end{center}
\caption{The behaviour of ${|\bar b_m({\infty})| \over m! }$. }
\label{b_bar}
\end{figure}

Next we investigate the effect of including all the  $m\geq 0$
terms in~(\ref{qt_sum1}). (To be approximated numerically the series has to be
truncated at some $N_{\rm max}$. Thus full evaluation of~(\ref{qt_sum1})
up to the $N_{\rm max}$-th term
requires knowledge of the first $2N_{\rm max}-1$ coefficients $\bar
b_m(\infty)$.)
The first 20 coefficients, calculated according to~(\ref{b_form}), are listed in Table~\ref{b_coeff}. 
We find (see Fig.~\ref{b_bar}) that for large $m$ the 
coefficients behave as \mbox{$ \bar{b}_m(\infty) \approx C\,(-1)^m m! 2^{-m}$}, where 
$C$ is a constant.

\begin{figure}[p]
\begin{center}
\mbox{\epsfig{figure=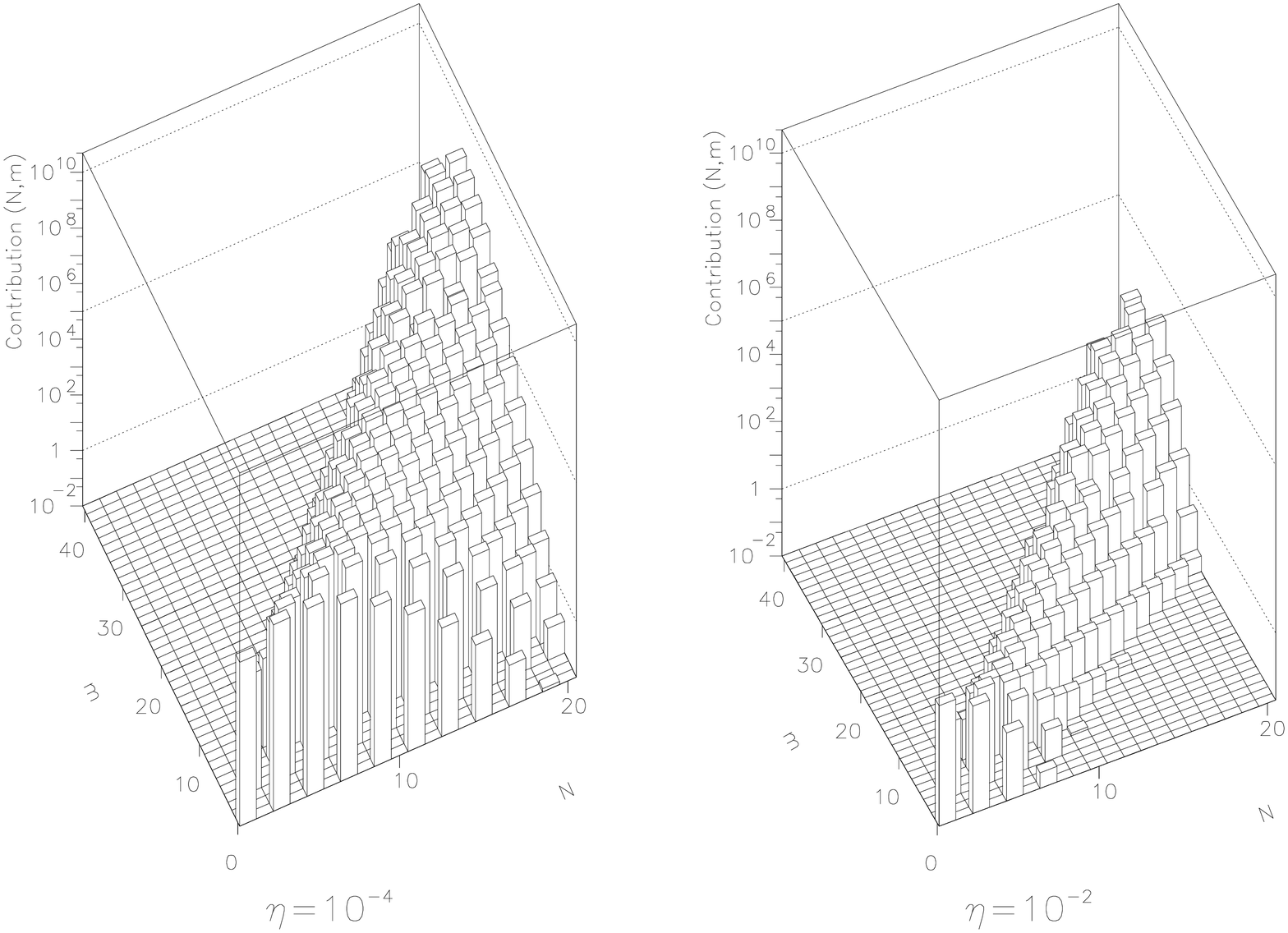,height=10cm,width=15cm}}
\end{center}
\caption{Contributions~(\ref{contrib_sum1}) to the
cross section~(\ref{qt_sum1}). Only positive contributions plotted here.}
\label{contrib:sum1}
\end{figure}

Taking more terms into account, i.e. $m\geq 1$, we find that, as expected, the
sum~(\ref{qt_sum1}) exhibits behaviour consistent with an asymptotic 
series.
A single $(N,m)$ contribution to~(\ref{qt_sum1}) is of the form   
\be
{1 \over \eta} \lambda^N { (-1/2)^{N-1} \over (N-1)!} 
{ 2^m \bar{b}_m(\infty) } { \left( \begin{array}{c} 2N-1 \\ m \end{array} \right)}
\ln^{2N-1-m}  \bigg ({1 \over \eta} \bigg)\; ,
\label{contrib_sum1}
\ee
and to show the complexity of the resummation~(\ref{qt_sum1}) we display
 these individual 
contributions in Fig.~\ref{contrib:sum1}. For all $\eta$ the biggest
contributions arise when $m \sim 2N-1$, since the coefficients $ \bar
b_m(\infty)$ are largest there. As
$\eta$ decreases, contributions with smaller $m$ become more
significant due to the terms $\ln ^{2N-m-1}\left( {1 \over \eta}
\right)$.

\begin{figure}[p]
\begin{center}
\mbox{\epsfig{figure=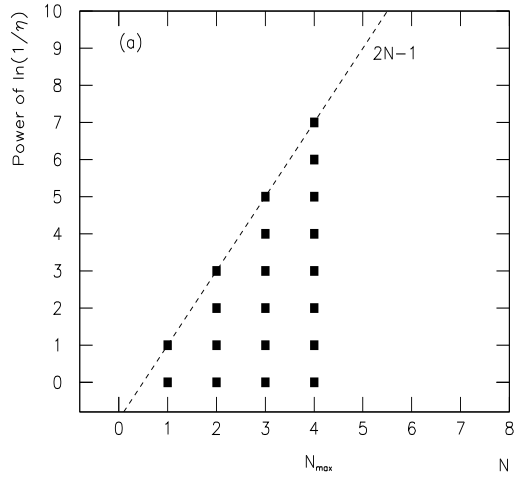,height=8cm,width=8cm}}
\mbox{\epsfig{figure=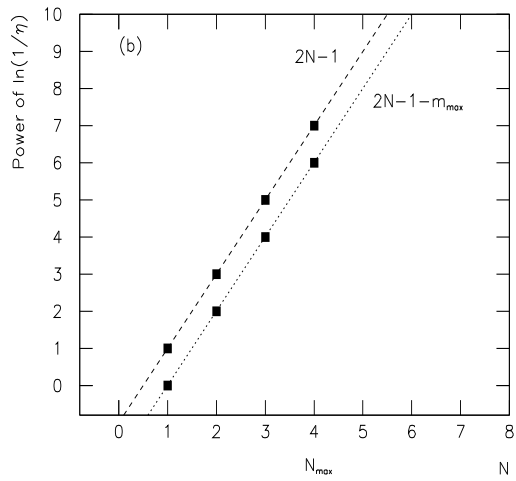,height=8cm,width=8cm}}
\end{center}
\caption{Resummation of~(\ref{contrib_sum1}).
Each point corresponds to a contribution~(\ref{contrib_sum1}) summed
in~(\ref{qt_sum1}) when (a): `all' $m_{max} \geq 2N_{\rm max}-1$ coefficients 
$\bar b_m({\infty})$ are known and (b): only $m_{max}<2N_{\rm max}-1$ are known.
In particular here $N_{\rm max}=4$ and $m_{\rm max}=7,1$ for the case (a),(b),
respectively. Here $N$ equals power of the coupling $\alpha_S$.} 
\label{draw1}
\end{figure}

\begin{figure}
\begin{center}
\mbox{\epsfig{figure=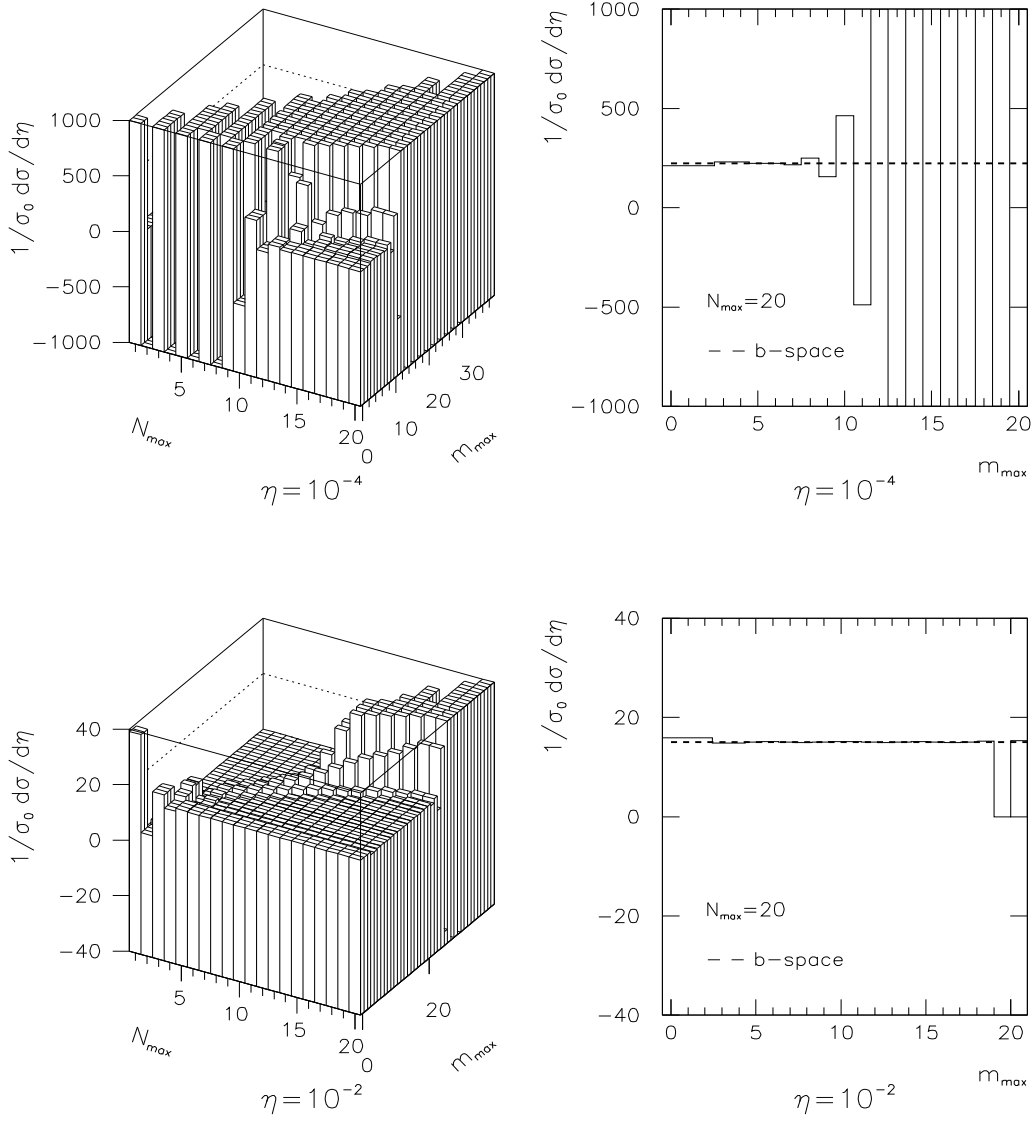,height=18cm,width=16cm}}
\end{center}
\caption{The cumulative plot of~(\ref{qt_sum1}) for
$\eta=10^{-4},10^{-2}$ and its section along $N_{\rm{max}}=20$.}
\label{3D_sum1}
\end{figure}

Our first task is to investigate numerically
the dependence of~(\ref{qt_sum1}) on the point of truncation
$N_{\rm max}$, i.e. the order of the perturbative expansion,
and the number of terms included in the internal
summation~(\ref{qt_sum1}) -- the `cut-off' value $m_{\rm max}$, 
equivalent to the  number of known `towers' of logs down from leading. 
Obviously for different pairs ($N_{\rm max}, m_{\rm max}$) different
contributions~(\ref{contrib_sum1}) are summed, see Fig.~\ref{draw1}. 
In Fig.~\ref{3D_sum1} we show a 3D cumulative plot of~(\ref{qt_sum1})
which illustrates some of the features discussed below. Each plotted
value for a given point $(N_{\rm max}, m_{\rm max})$ represents the 
sum~(\ref{qt_sum1}), truncated at $N_{\rm max}$ and calculated with
$\bar b_m(\infty)=0$ for $m >m_{\rm max}$. The distinctive plateau present
for large values of $N_{\rm max}$ and small $m_{\rm max}$ is equivalent to
recovering the $b$-space result for various $\eta$. Notice how for smaller $\eta$ this
plateau has a tendency to contract. If all $2N_{\rm max}-1= m_{\rm max}$
coefficients $\bar b_m(\infty)$ are taken into account (see Fig.~\ref{draw1}a), it seems
that the $b$-space result cannot be approximated for any value of $N_{\rm
  max}$, except for the region of large $\eta$. This should not be
surprising, considering that the `towers' of logarithms have been {\it truncated}.
Conversely, if only the first few coefficients ($m_{\rm max} < 2N_{\rm
  max}-1$) are known (Fig.~\ref{draw1}b) and the rest of them are set to zero, 
then in some sense one is closer to the DLLA situation and it is possible to
find $N_{\rm max}$ such that the cross
section~(\ref{qt_sum1}) approaches the $b$-space result, at least for the
range of $\eta$ considered here. Moreover, it can be seen from
Fig.~\ref{3D_sum1} that it is necessary to consider the first {\it few} coefficients 
to achieve the best approximation of the $b$-space result.

So far we have not attempted any analytic resummation of the series 
for the  \mbox{$q_T$-space} cross section given in (\ref{qt_sum1}). 
It is interesting to see whether
factorizing out the resummed DLLA piece from (\ref{qt_sum1})
 leads to an improvement in the approximation
of the $b$-space result.
We again start from~(\ref{qt_sum1}) but now we extract  the Sudakov factor $\exp
\left(-{\lambda \over 2} \ln^2  \left( {1 \over
\eta} \right) \right) $ from the sum to get
\ba
{1 \over \sigma_0} {d \sigma \over d \eta} &=& 
{\lambda \over \eta} e^{ {-\lambda \over 2}  \ln^2 \big({1 \over \eta}\big)}
\sum_{N=1}^{\infty} {(-2 \lambda)^{(N-1)} \over (N-1)!} 
\sum_{m=0}^{N-1} { \left( \begin{array}{c} N-1 \\ m \end{array} \right)}
\ln^{N-1-m}  \bigg( {1 \over \eta} \bigg)\nn \\
&\times& \bigg[
2\bar{b}_{N+m}(\infty)+ \ln \bigg({1 \over \eta} \bigg) \bar{b}_{N+m-1}(\infty)
 \bigg] \ .
\label{qt_sum2}
\ea
A key feature of (\ref{qt_sum2}) is that 
after extracting the Sudakov factor, the residual perturbation series has at most
$N+1$ logarithms of $1/\eta$ at $N$th order in perturbation theory, i.e. the leading terms
are now $\lambda^N \ln^{N+1}\left( {1 \over\eta} \right) $. However we know that
these terms must sum to give a large contribution as $\eta \to 0$ in order
to compensate the overall suppression from the Sudakov factor.

The terms which contribute to the new series~(\ref{qt_sum2}) are illustrated 
schematically in Fig.~\ref{draw2}. Notice that the extraction of the Sudakov
factor  results in an ability to sum an  {\it infinite} subseries of
logarithms. This observation constitutes a basis for our further analysis.
However, there is a shortcoming: in order to sum the first $m$ `towers' {\it fully}
we need to take $N_{\rm max}=m$ which leads us to include extra sub-leading
contributions from more than the first $m$ `towers', cf. Fig.~\ref{draw2}b.

\begin{figure}[p]
\begin{center}
\mbox{\epsfig{figure=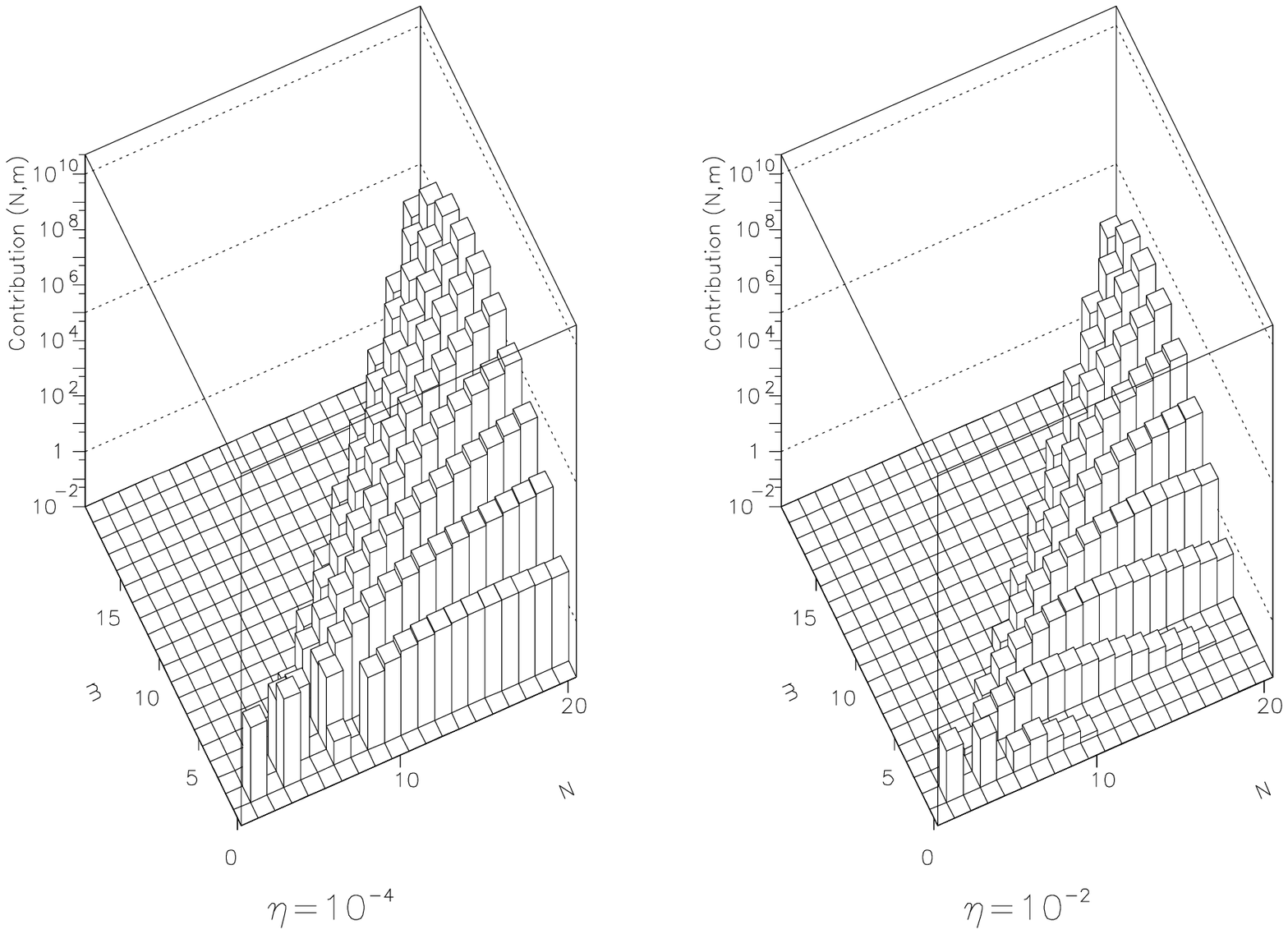,height=12cm,width=16cm}}
\end{center}
\caption{Contributions~(\ref{contrib_sum2}) to the
cross section~(\ref{qt_sum2}). Only positive contributions plotted here.}
\label{contrib:sum2}
\end{figure}

\begin{figure}[p]
\begin{center}
\mbox{\epsfig{figure=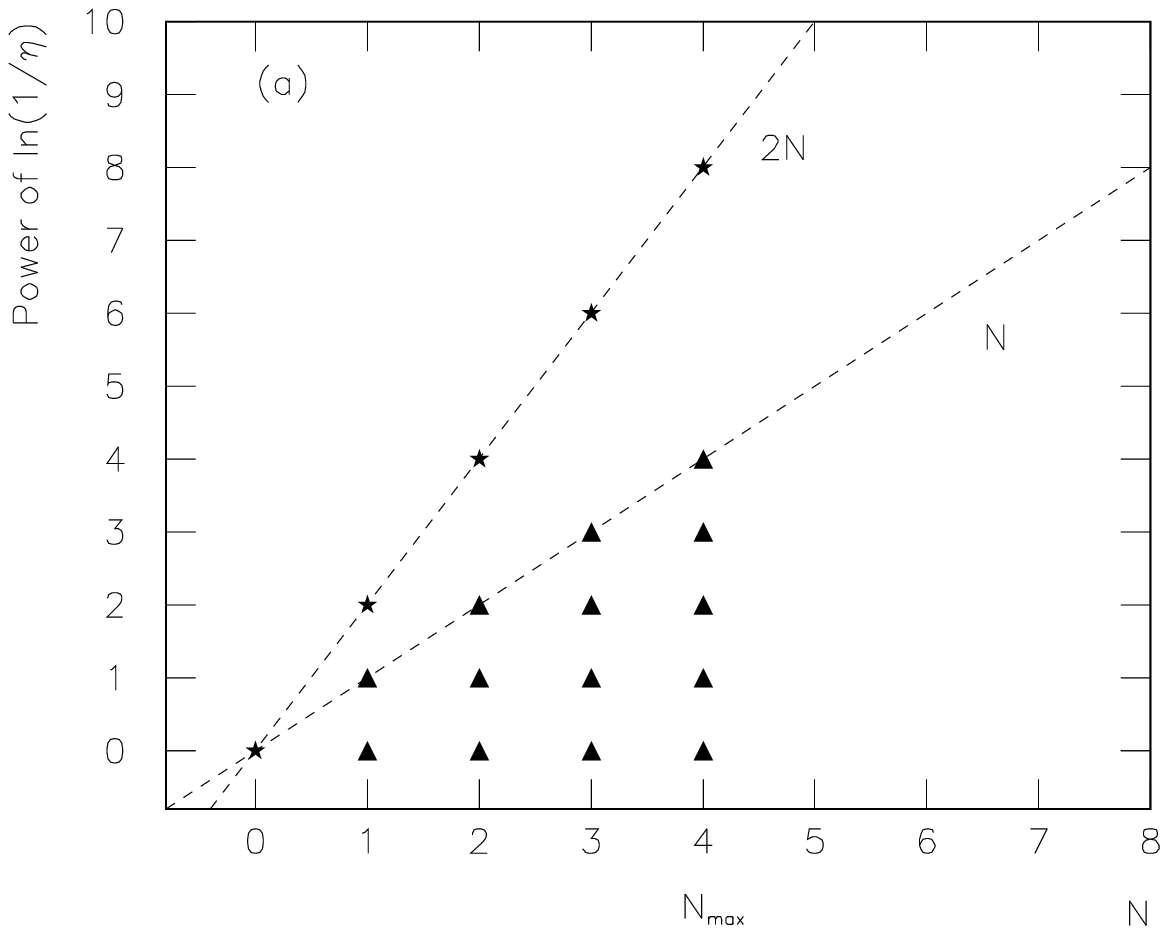,height=8cm,width=8cm}
\epsfig{figure=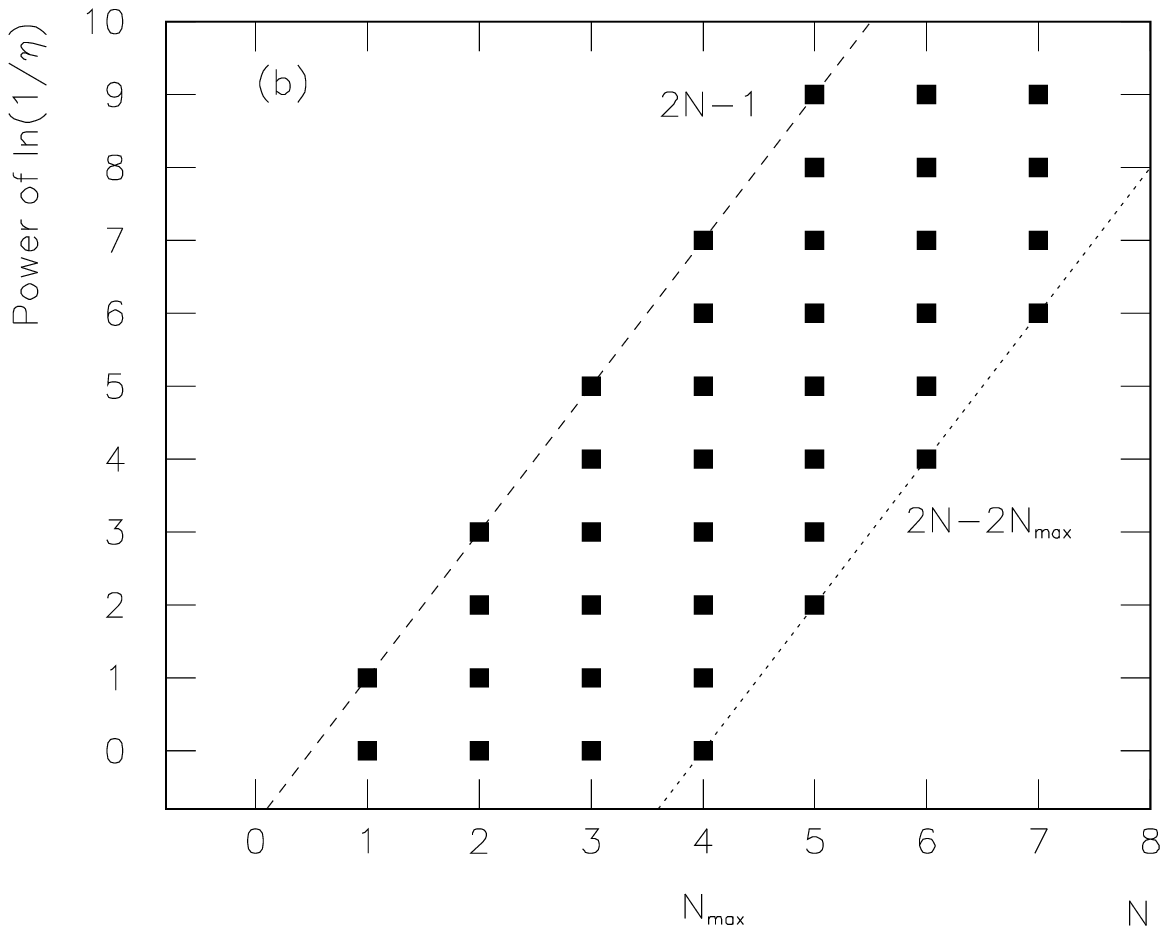,height=8cm,width=8cm}}
\mbox{\epsfig{figure=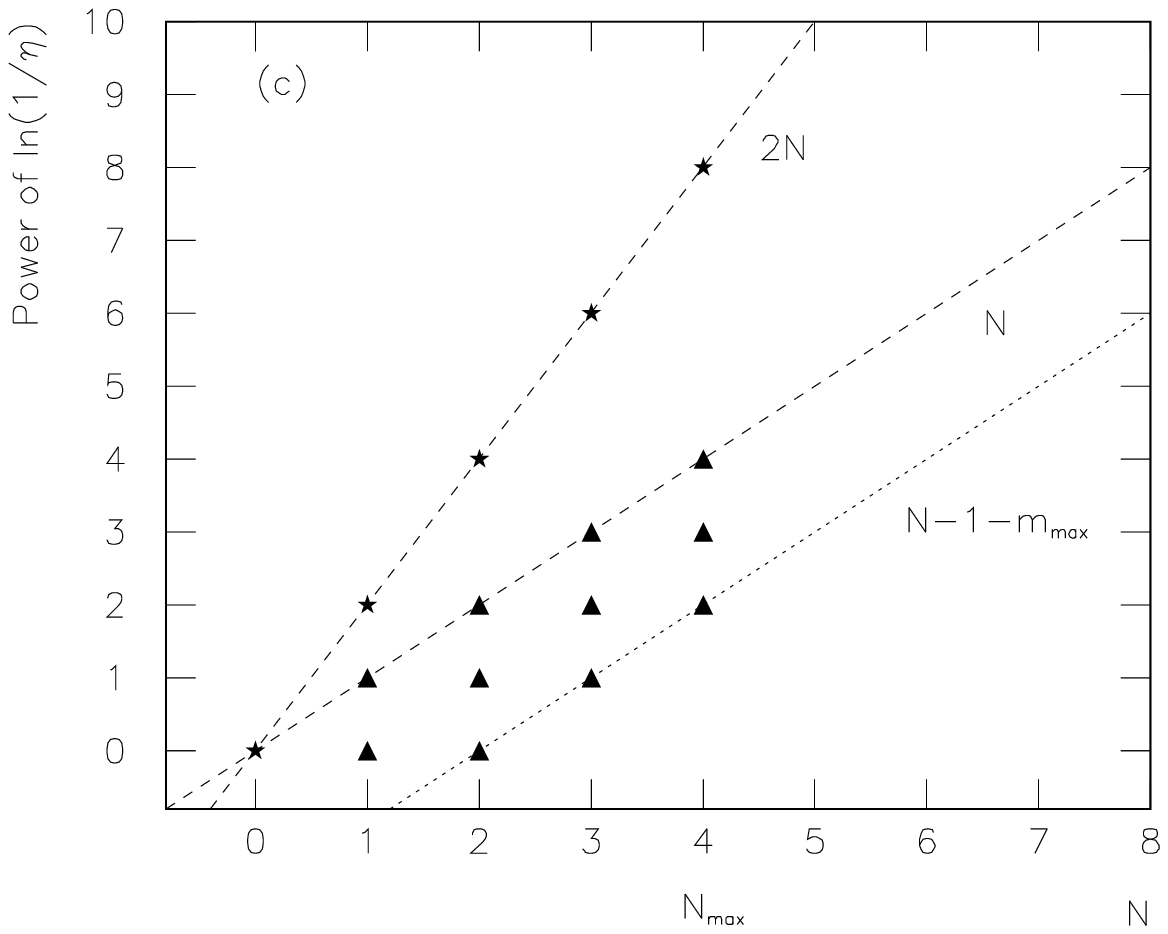,height=8cm,width=8cm}
\epsfig{figure=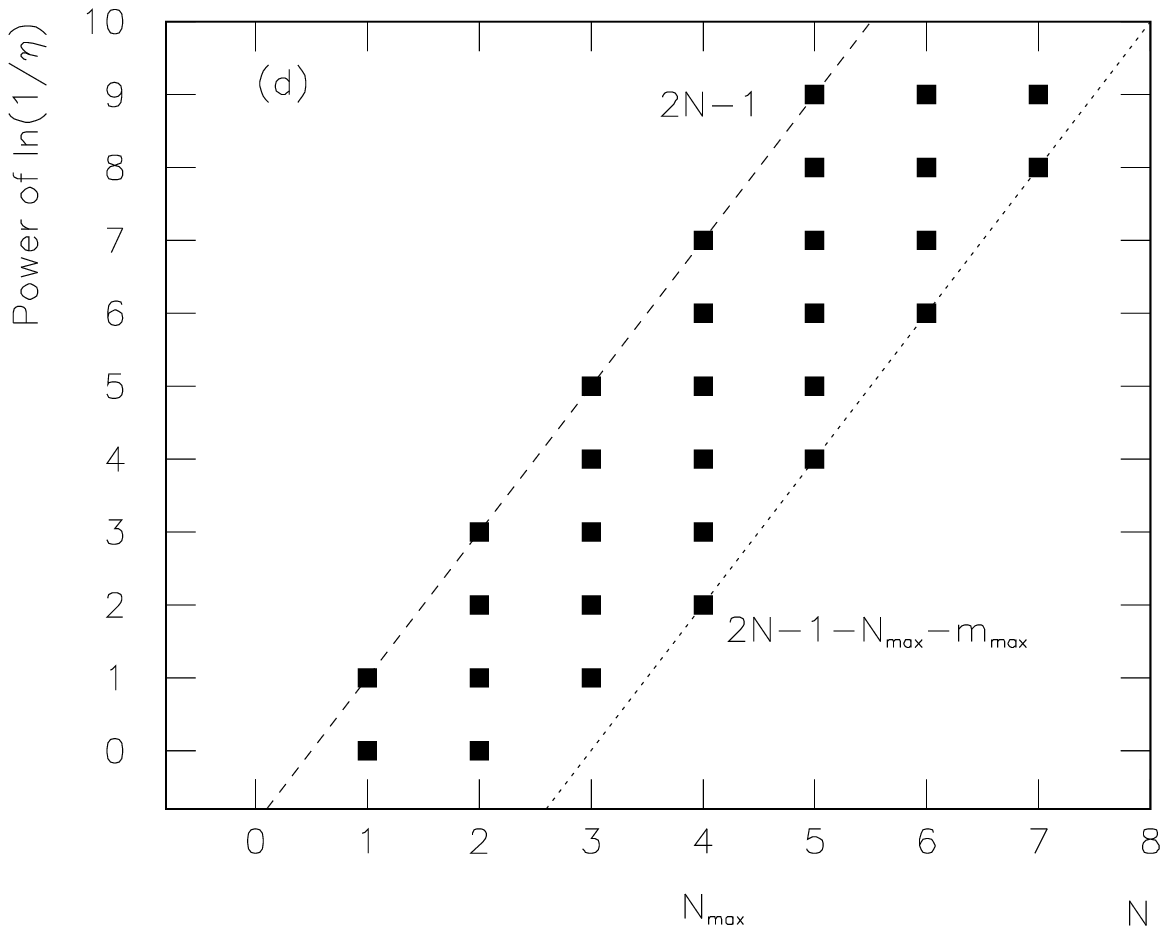,height=8cm,width=8cm}}
\end{center}
\caption{Resummation of Eq.~(\ref{contrib_sum2}).
Each point corresponds to a contribution~(\ref{contrib_sum2}). The
points  along the straight line `power of \mbox{$\ln \big( {1 \over \eta}
\big)=2N$'} represent terms coming from the Sudakov factor. Figures 
(b) and (d) illustrate contributions summed when this factor is expanded. 
In particular, here $N_{\rm max}=4$ and $m_{\rm max} = 7$ for the case
(a),(b) and $m_{\rm max}=1$ in (c),~(d). Note that only the 
$N_{\rm max}$, min$(N_{\rm max},2+2m_{\rm max})$ first `towers' are 
fully summed in (b),~(d), respectively. } 
\label{draw2}
\end{figure}

The individual contributions to the summation (\ref{qt_sum2}),
\ba
{1 \over \eta} e^{- {\lambda \over 2}  \ln^2 \big({1 \over \eta}
\big)}
{(-2)^{(N-1)} \lambda^N \over m! (N-1-m)!}
\ln  ^{N-1-m}\bigg( {1 \over \eta} \bigg) 
\bigg[
2\bar{b}_{N+m}(\infty)+ \ln \bigg({1 \over \eta} \bigg) \bar{b}_{N+m-1}(\infty)
 \bigg] \ ,
\label{contrib_sum2}
\ea
are presented in Fig.~\ref{contrib:sum2}. As was the case
for~(\ref{contrib_sum1}), the importance of the $\ln \left( {1 \over \eta} \right)$
factors diminishes for small $m$ and large $\eta$. Note that for small $\eta$
the sizes of the contributions are much smaller here than for~(\ref{qt_sum1}).
 
\begin{figure}[p]
\begin{center}
\mbox{\epsfig{figure=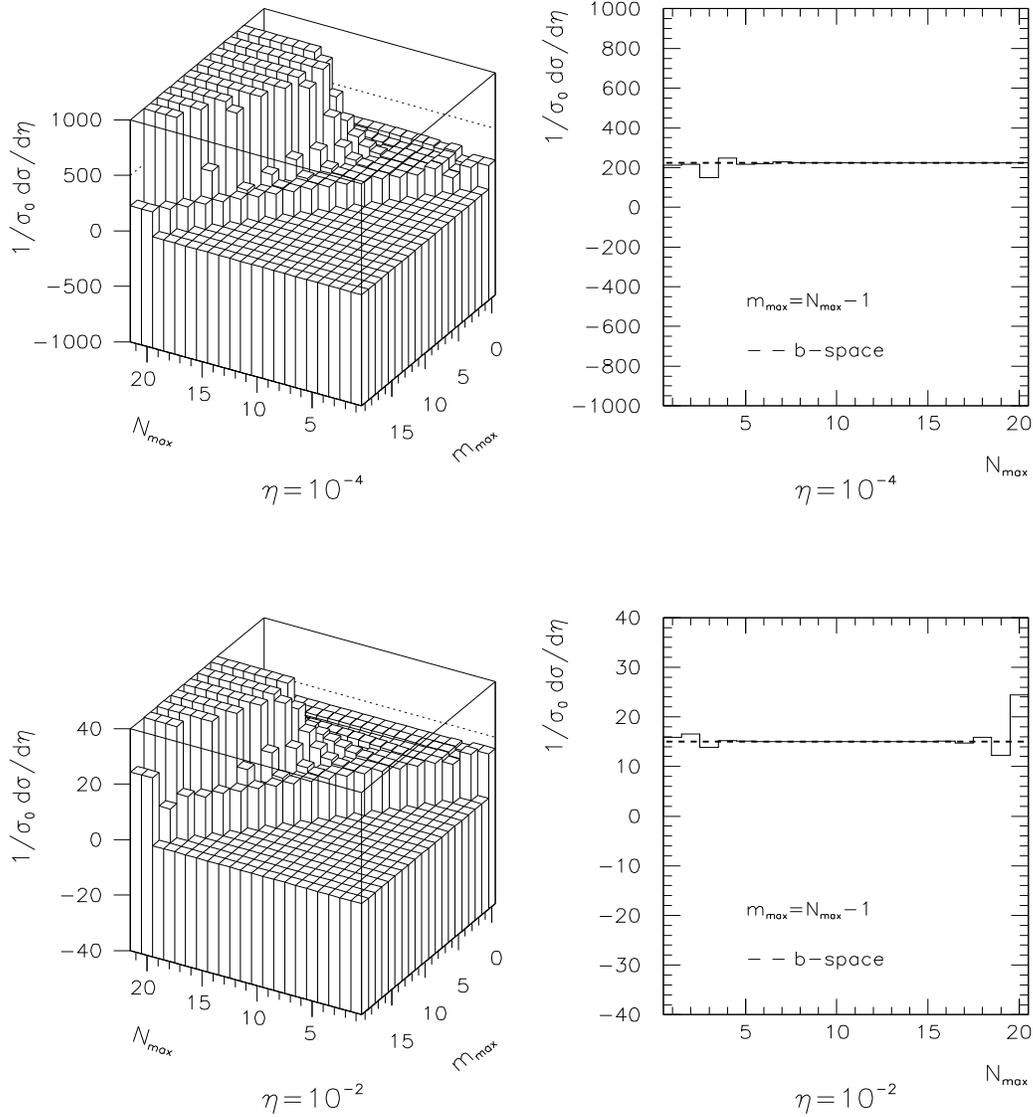,height=18cm,width=16cm}}
\end{center}
\caption{The cumulative plot of~(\ref{qt_sum2}) for
$\eta=10^{-4},10^{-2}$ and its section through the line $m_{\rm max}=N_{\rm max}-1$.}
\label{3D_sum2}
\end{figure}

Next we perform the same analysis as for~(\ref{qt_sum1}), i.e. we study the
dependence of~(\ref{qt_sum2}) on $N_{\rm max}$ and on the `cut-off' value
$m_{\rm max}$. 
Comparing the cumulative plot for~(\ref{qt_sum2}), Fig.~\ref{3D_sum2}, with
Fig.~\ref{3D_sum1}, we note that the $b$-space result is now better approached
close to the line $m_{\rm max}=N_{\rm max}-1$ (such an effect can be observed 
in the case of~(\ref{qt_sum1}) only for large $\eta$). When $m_{\rm max} < N_{\rm
  max}-1$  the balance between different contributions is apparently 
spoiled until $m_{\rm max}$ becomes considerably small. Again, it turns out that
it is necessary to incorporate the first {\it
  few} sub-leading kinematic logarithmic terms (i.e. some moderate $N_{\rm max}$) to obtain the best approximation of the $b$-space result.

The asymptotic properties of~(\ref{qt_sum2}) can
be most easily seen for large values of $\eta$ and large $N_{\rm max}$. The apparent
discrepancy between the $b$-space result and the summation~(\ref{qt_sum2}) is
caused here by contributions with $m=N-1$, i.e. terms proportional to  $\bar
b_{2N-1}(\infty)$. The other terms are negligible as they contain
powers of $\ln \left( {1 \over \eta} \right)$ which for these $\eta$
are very small. The dominant contribution is then proportional to
$(-1)^{N_{\rm max}-1}2^{N_{\rm max}}\bar{b}_{2N_{\rm max}-1}(\infty)/(N_{\rm
  max}-1)!$, i.e. the sign varies as $(-1)^{N_{\rm max}}$.
Fortunately, the range of $\eta$ for which the discrepancy occurs is
outside the region of interest for the present discussion. Also, in practice
we never take $N_{\rm max}$ so large as to make this effect substantial.
Nevertheless, it  emphasises  the necessity of performing a careful
matching with the fixed-order perturbative result.

We may therefore conclude that the expression~(\ref{qt_sum2}), with the Sudakov factor resummed
and factored out, enables us to resum infinite series of logarithms while at the
same time allows for the inclusion of kinematical sub-leading logarithms. Moreover, it 
has very good convergence
properties over a  large range of $\eta$, while summing leading and
sub-leading logarithmic terms. It is thus well suited for the
purposes of this analysis, i.e. reproducing the $b$-space result
by explicit resummation in $q_T$ space,  and we will continue to use it for the rest of this study.

\subsection{Running coupling analysis}

\begin{figure}
\begin{center}
\mbox{\epsfig{figure=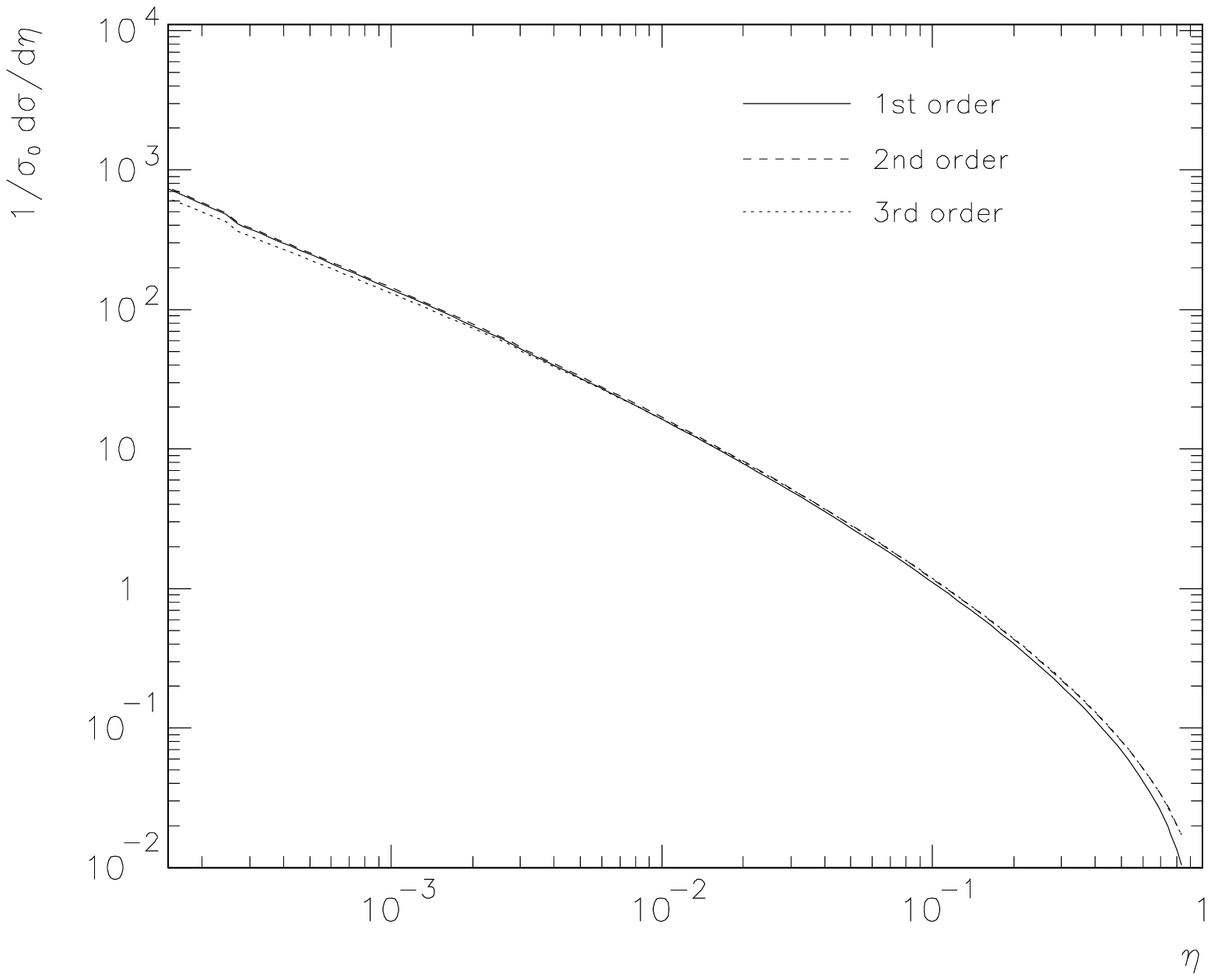,height=9cm,width=14cm}} 
\vspace{1.5cm}
\mbox{\epsfig{figure=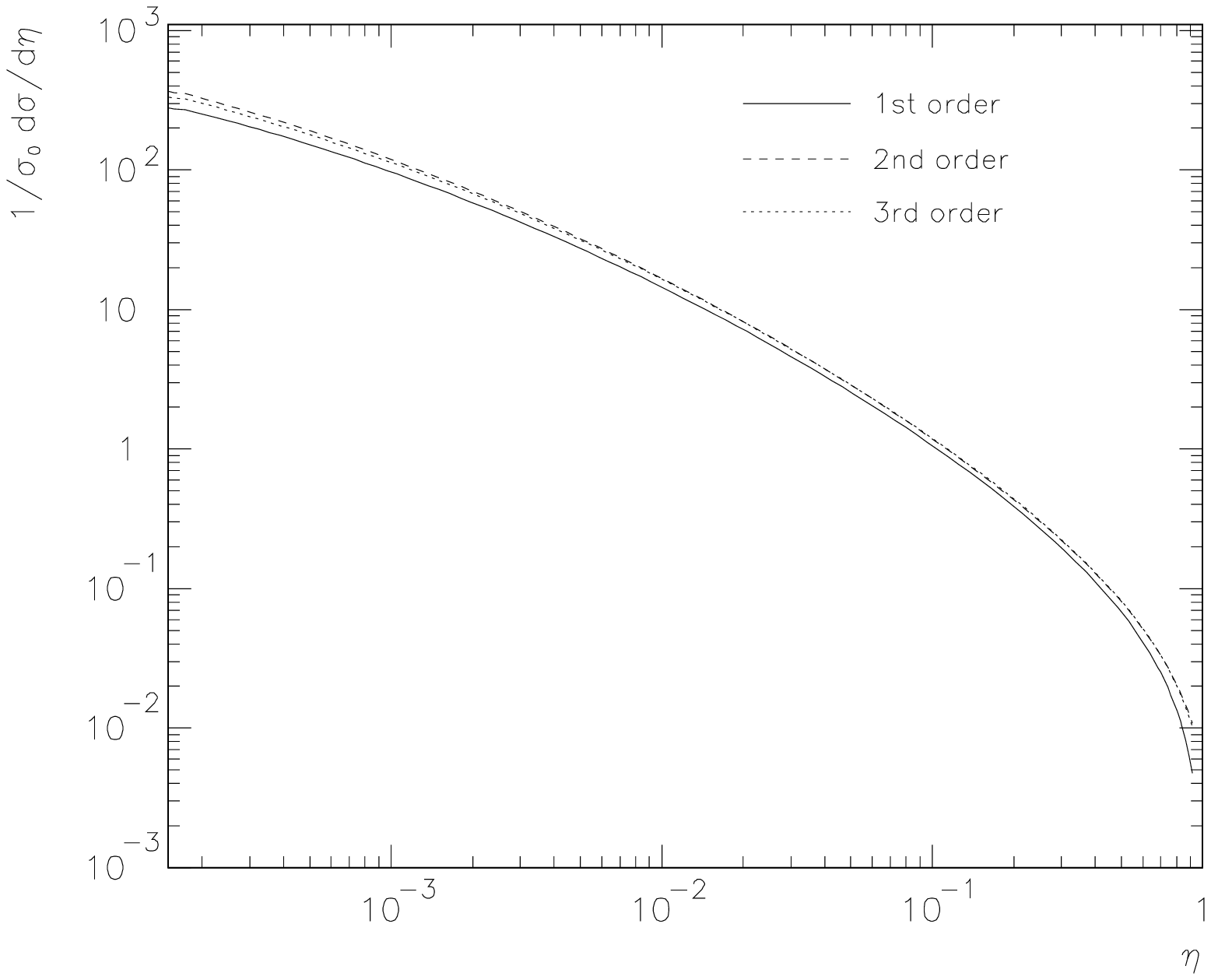,height=9cm,width=14cm}}
\end{center}
\vspace{-2cm}
\caption{Comparison between the resummed~(\ref{beta0}) using two different choices
of renormalization scale: $\mu^2=q_T^2$ (top figure) and 
$\mu^2=Q^{2 /3 }q_T^{4 /3 }$  (bottom figure), and the first three orders in
$\alpha_S(\mu^2)$ in~(\ref{beta0}) (i.e. the orders in  $\alpha_S(\mu^2)$ at
which the residual sum is truncated). Here $Q=M_Z=91.187$ GeV, $\alpha_S(M_Z^2)=0.113$.}
\label{mu2_depend}
\end{figure}

The ultimate goal of the work presented here is to obtain a more
accurate description, if possible, of the $W$ production distribution. 
To this end, one  has to incorporate various other sub-leading effects in 
addition
to the kinematic logarithms discussed in the previous subsection. One example
is the incorporation of the running of the strong coupling
$\alpha_S$ into the cross section expression. This is achieved by substituting
\ba
\alpha_S(\bar\mu^2) = \alpha_S(\mu^2) &\bigg\{& \!\! 1 -  {\alpha_S(\mu^2)\over 
\pi}\beta_0 \ln({\bar\mu^2 \over \mu^2}) \nn \\
&& + {\alpha_S(\mu^2)^2 \over \pi^2}
\ln({\bar\mu^2 \over \mu^2}) \left( \beta_0^2 \ln({\bar\mu^2 \over \mu^2})
  -\beta_1 \right) \bigg\} + {\cal O}(\alpha_S^3(\mu^2)) \,,
\label{runningas}
\ea
with
\be
\beta_0= {1 \over 4 }\left ( 11 -{2 \over 3} n_f \right)\,, \qquad
\beta_1= {1 \over 16 }\left ( 102 -{38 \over 3} n_f \right)\,,
\ee
in (\ref{qt_sum2}). The effect on the DLLA form factor is to introduce
a sequence of sub-leading logarithms whose coefficients depend
on the $\beta$--function coefficients defined in (\ref{runningas}).
If only the 1-loop running of $\alpha_S$ is introduced
then the cross section expression (\ref{qt_sum2}) reads
\ba
&{1 \over \sigma_0} {d \sigma \over d \eta}& = {\alpha_S(\mu^2)A^{(1)} \over
   2\pi\eta}
e^{ -{\alpha_S(\mu^2)A^{(1)} \over 4 \pi}\ln^2\big( {1 \over \eta}\big) \big[ 1 + 
{1 \over 3 \pi} \beta_0\alpha_S(\mu^2) 
 \big( 2 \ln  \big( {1 \over \eta}\big) - 3 \ln \big({Q^2\over\mu^2}\big) \big)\big]}
\nn \\ && \times 
\sum_{N=1}^{\infty} \Bigg(- {\alpha_S(\mu^2)A^{(1)} \over \pi}\Bigg)^{N-1} { 1\over (N-1)!} 
\sum_{m=0}^{N-1} { \left( \begin{array}{c} N-1 \\ m \end{array} \right)}
\sum_{k=0}^{N-m-1}{ \left( \begin{array}{c} N-m-1 \\ k \end{array} \right)}
\nn \\
&&c_2^m c_3^k c_1^{N-m-k-1} \times \bigg[
3 c_3 \bar{b}_{N+m+2k+1}(\infty) + 2 c_2 \bar{b}_{N+m+2k}(\infty)
+ c_1 \bar{b}_{N+m+2k-1}(\infty)
 \bigg] \,,\qquad
\label{beta0}
\ea
with  
\ba
&&c_3= {4 \over 3\pi} \beta_0 \alpha_S(\mu^2) \,, \nn \\
&&c_2= {1 \over \pi} \bigg[ \pi + \beta_0\alpha_S(\mu^2)
\bigg( 2\ln \bigg( {1 \over\eta}\bigg) - \ln \bigg({Q^2\over\mu^2}\bigg)\bigg)
\bigg] \,,  \nn \\
&&c_1= {1 \over \pi}\ln \bigg( {1 \over\eta}\bigg) 
\bigg[ \pi + \beta_0\alpha_S(\mu^2)
\bigg( \ln \bigg( {1 \over\eta}\bigg) - \ln \bigg({Q^2\over\mu^2}\bigg) 
\bigg)\bigg]\,. 
\label{ccoeffs}
\ea
Notice the appearance in (\ref{beta0}) of sub-leading ${\cal O}(\alpha_S^2 \ln^3(1/\eta))$
terms in the exponent of the Sudakov form factor. Interestingly, these
logarithms can be eliminated by a particular scale choice: 
$\mu^2= Q^2\,\eta^{2/3}$. This restores 
the same form as in~(\ref{qt_sum2}) i.e.
$\exp{ \big(- {\alpha_S A^{(1)}\over \pi}
  \ln^2 \big( {1 \over \eta} \big) \big)}$, but now with a coupling which also
depends on $\eta$.%
\footnote{There is an  analogous $y_{\rm cut}$--dependent
scale choice for resummed jet cross sections
in $e^+e^-$ annihilation, see for example~\cite{Brown:Stirling}. }
However not all $\beta_0$--dependent
logarithms are eliminated. For example, we can see from 
(\ref{ccoeffs}) that corrections of order
$\beta_0 \alpha_S\ln \big( {1 \over \eta} \big)$ remain.
Of course in a complete calculation the dependence on the scale choice
should disappear. To illustrate the residual dependence on the scale
of the cross section (\ref{beta0})  we show (Fig.~\ref{mu2_depend})
 results for two different
choices: $\mu^2=\eta\,Q^2= q_T^2$ and
$\mu^2=Q^2\,\eta^{2/3} $. Also shown is the effect of
truncating the sum in (\ref{beta0}) at first, second and third order.%
{\footnote{Obviously, now we truncate the sum in~(\ref{beta0}) at a certain
   order of $\alpha_S(\mu^2)$, so that `$N$' in  Fig.~\ref{draw2} should be
   read as `power of $\alpha_S$'.}

We see that with $\mu^2 =Q^2\,\eta^{2/3} $ there is slightly less stability with respect to the truncation point than there is with the choice $\mu^2=q_T^2 $.
Note also  that now the plots 
of ${1 \over \sigma_0} { d\sigma \over d\eta}$ are no  longer
 explicitly independent of $Q$, as  was the case when $\alpha_S$ was taken to be constant.
The additional dependence
on $Q$ enters into~(\ref{beta0}) via the coupling
$\alpha_S(\mu^2)=\alpha_S(Q^{2/3} q_T^{ 4/3})$. For values of $q_T$
where perturbative QCD can safely  be applied (e.g. $q_T\geq 3$ GeV) and the
energies considered here, the resulting scale $\mu$ is bigger than the 
$b$ quark mass, and  we therefore avoid evolving the coupling through any quark 
mass thresholds. 

An extension of this analysis for the case of the two-loop running coupling is
straightforward and gives  
\ba
&{1 \over \sigma_0} {d \sigma \over d \eta}& = {\alpha_S(\mu^2) A^{(1)}\over 2\pi\eta}
e^{ S_{\eta}} \sum_{N=1}^{\infty} \Bigg(- { \alpha_S(\mu^2) A^{(1)}\over \pi}\Bigg)^{N-1} { 
  1 \over (N-1)!}
\sum_{m=0}^{N-1} { \left( \begin{array}{c} N-1 \\ m \end{array} \right)}
\nn \\  
&& \times
\sum_{k=0}^{N-m-1}{ \left( \begin{array}{c} N-m-1 \\ k \end{array} \right)}
\sum_{l=0}^{N-m-k-1}{ \left( \begin{array}{c} N-m-k-1 \\ l \end{array} \right)}
c_2^m c_3^k c_4^l c_1^{N-m-k-l-1} \nn\\
&&\times \bigg[
4 c_4 \bar{b}_{N+m+2k+3l+2}(\infty) + 3 c_3 \bar{b}_{N+m+2k+3l+1}(\infty) 
+ 2 c_2 \bar{b}_{N+m+2k+3l}(\infty)\nn\\
&& \ + c_1 \bar{b}_{N+m+2k+3l-1}(\infty)
 \bigg] \,.
\label{beta1}
\ea
Here
\bann
S_{\eta}&=& - {\alpha_S(\mu^2) A^{(1)} \over 4\pi} \ln^2 \bigg( {1 \over\eta}\bigg) \bigg[
1 +   {1 \over 3\pi}\bigg(\beta_0\alpha_S(\mu^2) +  {1 \over\pi}
\beta_1\alpha_S^2(\mu^2) \bigg) \bigg( 2 \ln  \bigg( {1 \over \eta}\bigg) - 3
\ln \bigg({Q^2\over\mu^2}\bigg) \bigg) \\
&& + {1 \over 6\pi^2}\alpha_S^2(\mu^2)
\beta_0^2 \bigg(3\ln^2 \bigg( {1 \over\eta}\bigg) -8 \ln \bigg( {1\over\eta}
\bigg) \ln \bigg({Q^2\over\mu^2}\bigg) + 6\ln^2
\bigg({Q^2\over\mu^2}\bigg)\bigg)\bigg] \,,  \\
%%%
c_4&=&{2 \over \pi^2}\alpha_S^2(\mu^2)\beta_0^2\,, \\
%%%
c_3&=&{4 \over 3\pi^2}\bigg[  \beta_0 \alpha_S(\mu^2)\pi +
\beta_1\alpha_S^2(\mu^2) +
 \beta_0^2 \alpha_S^2(\mu^2) \bigg( 3 \ln  \bigg( {1
  \over \eta}\bigg) - 2\ln \bigg({Q^2\over\mu^2}\bigg) \bigg) \bigg] \,,\\ 
%%%
c_2&=& {1 \over \pi^2} \bigg[ \pi^2 + \bigg(\beta_0\alpha_S(\mu^2)\pi
+ \beta_1\alpha_S^2(\mu^2)\bigg)   
\bigg( 2\ln \bigg( {1 \over\eta}\bigg) - \ln \bigg({Q^2\over\mu^2}\bigg)\bigg) 
\\
&&+\beta_0^2 \alpha_S^2(\mu^2)\bigg(3\ln^2 \bigg( {1 \over\eta}\bigg)-4 \ln
\bigg( {1\over\eta} \bigg) \ln \bigg({Q^2\over\mu^2}\bigg) + \ln^2
\bigg({Q^2\over\mu^2}\bigg)\bigg) \bigg] \,, \\
%%%
c_1&=& {1 \over \pi^2}\ln \bigg( {1 \over\eta}\bigg) 
\bigg[ \pi^2 + \bigg(\beta_0\alpha_S(\mu^2)\pi+\beta_1\alpha_S^2(\mu^2)\bigg)
\bigg( \ln \bigg( {1 \over\eta}\bigg) - \ln \bigg({Q^2\over\mu^2}\bigg)\bigg)
 \\
&&+ \beta_0^2 \alpha_S^2(\mu^2) \bigg(\ln^2 \bigg( {1 \over\eta}\bigg)-2 \ln
\bigg( {1\over\eta} \bigg) \ln \bigg({Q^2\over\mu^2}\bigg) + \ln^2 \bigg({Q^2\over\mu^2}\bigg)\bigg) \bigg]  \,.\\
\eann
Now, after fixing the choice of scale as described above, 
a new term proportional to
$\beta_0^2\alpha_S^3(\mu^2)\ln^4\big( {1 \over \eta}
\big)$ appears in the exponential.  As a result, the
  line of points corresponding to terms of the form $\alpha_S^N \ln^{2N-1}\big(
  {1 \over\eta}\big)$ in Fig.~\ref{draw2}a will now change to the set of points
  corresponding to $\alpha_S^{3N} \ln^{4N}\big({1 \over\eta}\big)$ or higher
  powers of $ \ln \big( {1 \over\eta}\big)$.}
Unlike the 
$\ln^3\big( {1 \over \eta} \big)$ terms, those with $\ln^4\big( {1 \over \eta}
\big)$ do {\it not} cancel when $\mu^2= Q^{2/3}
q_T^{ 4 /3}$ is chosen. It is impossible to cancel these
terms by {\it any} choice of the renormalization scale. 
On the other hand, the choice which
eliminates those  terms with $\ln^3\big( {1 \over \eta} \big)$ is also the
one that minimizes the coefficient of the
 $\ln^4\big( {1 \over \eta} \big)$ terms.

%%%%%%%%%%%%

\subsection{Resummation including all types of sub-leading logarithms}

A derivation of the expression for the cross section for the case when all
known leading and sub-leading 
coefficients, i.e. $A^{(1)},B^{(1)},A^{(2)},B^{(2)}$, are taken into
account follows the method introduced above and gives (fixed coupling
to begin with)
\ba
&{1 \over \sigma_0} {d \sigma \over d \eta}& = {\alpha_S A^{(1)} \over 2\pi\eta} 
e^{ S_{\eta}} \sum_{N=1}^{\infty} {\Bigg(-{\alpha_S A^{(1)} \over \pi}\Bigg)^{N-1}}{1 \over (N-1)!}
\sum_{m=0}^{N-1} {\left( \begin{array}{c} N-1 \\ m \end{array} \right)} \nn \\
&& \times
c_2^m c_1^{N-m-1} \bigg[ 2c_2 \bar{b}_{N+m}(\infty)
 + c_1 \bar{b}_{N+m-1}(\infty) \bigg] \,,
\label{sublead}
\ea
with
\bann
S_{\eta}&=& - {\alpha_S A^{(1)}\over 4\pi}\bigg[ \ln^2 \bigg( {1 \over\eta}\bigg) \bigg(
1 +   {\alpha_S \over 2 \pi}{A^{(2)} \over A^{(1)} }\bigg)+ \ln  \bigg( {1
  \over \eta}\bigg) \bigg( 2 {B^{(1)} \over A^{(1)}} + {\alpha_S \over \pi}
{B^{(2)} \over A^{(1)}} \bigg) \bigg] \,,\\
c_1&=& \bigg( 1 + {\alpha_S \over 2 \pi}{A^{(2)} \over A^{(1)} }\bigg) \ln
\bigg( {1 \over \eta}\bigg) + {B^{(1)} \over A^{(1)}} + {\alpha_S \over 2 \pi}
{B^{(2)} \over A^{(1)}}\,, \\
c_2&=& 1 +  {\alpha_S \over 2\pi}{A^{(2)} \over A^{(1)} } \,.\\
\eann
Now each logarithmic term in
the sum~(\ref{sublead}) acquires a factor which 
is a combination of the 
$A$s, $B$s and $\bar b(\infty)$'s. Notice that although
the various sub-leading logarithms are mixed together, they have 
a distinctive origin.
We have mentioned already that the DLLA (i.e. retaining only terms
of the form $A^{(1)}\alpha_S^N\ln^{2N-1}\big( {1 \over \eta} \big)$ )
corresponds to the situation where all gluons 
are soft and collinear and where strong ordering
of the transverse momenta and energies is imposed.
 We also know that other terms with
$A^{(1)}$ multiplied by the 
$\bar b_m(\infty)$
arise from using the soft and collinear approximation for the matrix element
but relaxing the strong-ordering condition.
The sub-leading terms with $B^{(1)}$ etc. correspond to the situation where at least one
gluon is either non-collinear or energetic.

In addition, extracting a Sudakov form
factor from the sum~(\ref{sublead}) \lq squeezes' it down to a summation
over $m$ from 0
to $N-1$, thereby reducing the number of fully known `towers' of logarithms: from the
first four to the first two. This can be appreciated by comparing the 
logarithmic coefficients appearing in  the expansion of
the cross section up to and including
the first three orders in $\alpha_S$ 
with (Table~\ref{tow:we}) and without (Table~\ref{tow:ellis}) 
the Sudakov form factor extracted.

  These tables also reveal another 
relevant property: sub-leading
coefficients like $A^{(2)}$ are associated
with $\bar b(\infty)$ factors whose indices are not as high as those 
 which accompany $A^{(1)}$. Both $\ln{\big( {1 \over \eta}\big)}$ terms
and $\bar b(\infty)$ factors  originate from the same
$\ln \big( {Q^2 b^2 \over b_0^2} \big)$ terms appearing at the very early stage
of the derivation. In particular, let us focus on a term with a particular power of $\alpha_S$
and  $\ln{\big( {1 \over \eta}\big)}$. To reproduce such a term one has to
take $\ln \big( {Q^2 b^2 \over b_0^2} \big)$ up to the
 appropriate power, depending 
on its associated coefficient (i.e. $A$ or $B$). For sub-leading
coefficients this power will be obviously lower than for more leading
ones. Hence the indices of the corresponding $\bar b(\infty)$ factors are lower for
more sub-leading coefficients. For example, the index of $\bar
b(\infty)$ accompanying  the first sub-leading unknown coefficient
$A^{(3)}$  would be at least four less than  
the index of a
corresponding $\bar b(\infty)$ for the $A^{(1)}$ coefficient, for given
powers of $\alpha_S$ and $\ln{\big( {1 \over \eta}\big)}$.
This observation provides us with a strong argument for justifying the
inclusion of known {\it parts} of logarithmic terms from sub-leading `towers' lower
than NNNL. Physically, the $\bar b(\infty)$ factors are of  kinematic origin
and as such they are much more relevant to the cross section
than perturbative lower order
coefficients in the expansion of $A(\alpha_S(\mu^2))$ and
$B(\alpha_S(\mu^2))$. While, as we have shown earlier,  resummation of the terms 
containing $\bar b(\infty)$ factors with higher indices can still be of 
numerical  importance for
the final result, terms with unknown higher-order perturbative coefficients
seem to contribute corrections of much smaller size. 

\begin{figure}
\begin{center}
\mbox{\epsfig{figure=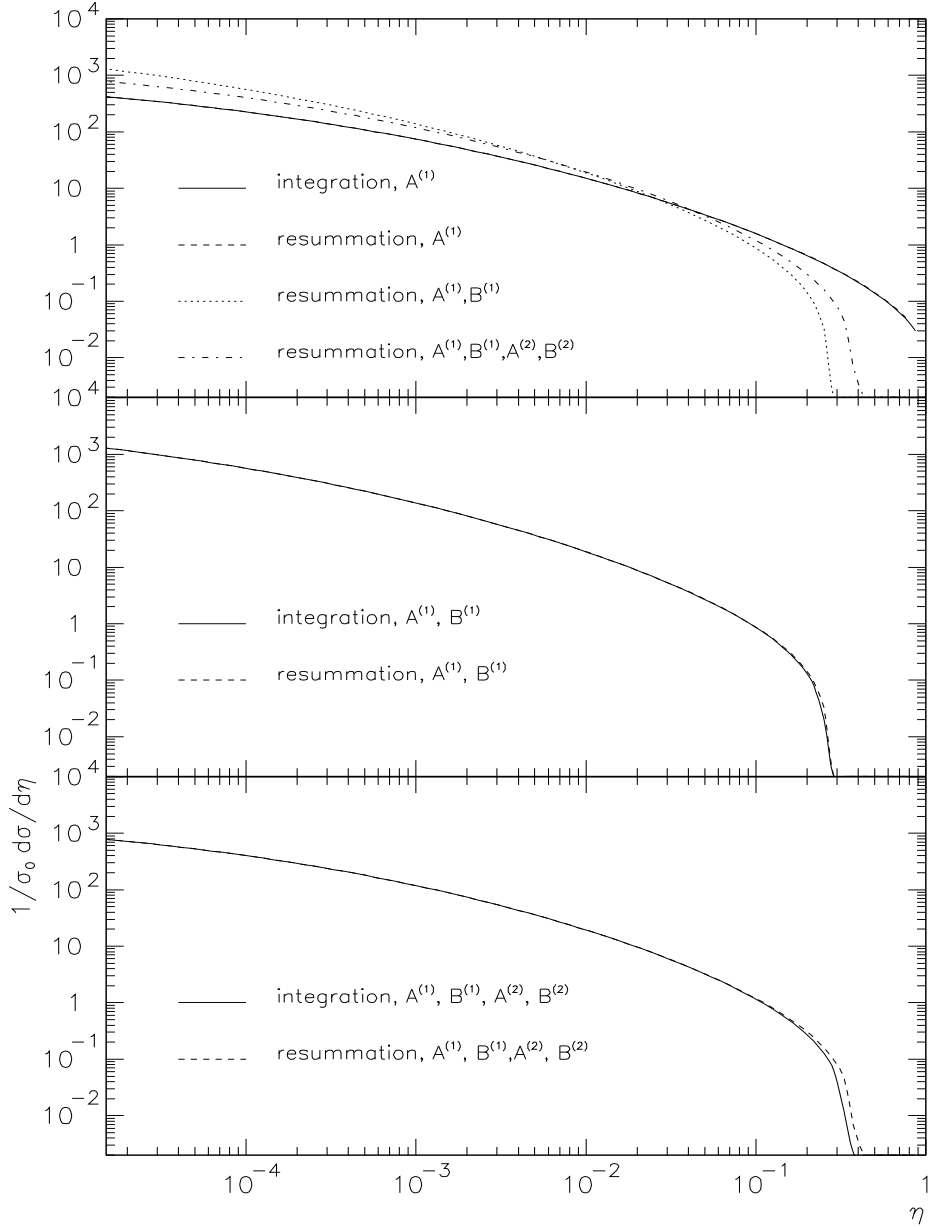,height=18cm,width=14cm}} 
\end{center}
\caption{The resummed cross section~(\ref{sublead}), truncated at
   $N_{\rm max}=10$, compared to the results of integration. Top figure: only
   $A^{(1)}$ coefficient non-zero, middle figure: $A^{(1)},B^{(1)}$ non-zero,
   bottom figure: $A^{(1)},B^{(1)}, A^{(2)},B^{(2)}$ non-zero. The top
   figure also shows a comparison of the  effects induced when subsequent coefficients are
   introduced.}
\label{subleadfig}
\end{figure}

A comparison of effects induced by the inclusion of successive sub-leading
coefficients in~(\ref{sublead}) is demonstrated in Fig.~\ref{subleadfig}. 
Also shown are the results of an exact  integration in $b$-space. The
agreement for low values of $\eta$ is excellent. At high
$\eta$  we encounter a discrepancy of the same nature as that discussed for
the case of the leading coefficient $A^{(1)}$,  
i.e. for large $N_{\rm max}$ the expression~(\ref{sublead})
either grows significantly or acquires negative values, depending on the
number of terms at which the expression is truncated. For the values of
$N_{\rm max}$ we use in our calculations this behaviour does not
occur.\footnote{This is also true for the case of the running coupling.}

With the above  prescriptions for the  sub-leading logarithmic terms,
kinematic effects and running coupling, we finally obtain
a `complete' expression for the cross section: 
\ba
&{1 \over \sigma_0} {d \sigma \over d \eta}& = {\alpha_S(\mu^2) A^{(1)} \over 2\eta\pi} 
e^{ S_{\eta}} \sum_{N=1}^{\infty} \Bigg({-\alpha_S(\mu^2) A^{(1)}\over
  \pi}\Bigg)^{N-1}{1  \over (N-1)!}
\sum_{m=0}^{N-1} { \left( \begin{array}{c} N-1 \\ m \end{array} \right)} \nn
\\
&\times&
\sum_{k=0}^{N-m-1} { \left( \begin{array}{c} N-m-1 \\ k \end{array} \right)}
\sum_{l=0}^{N-m-k-1} { \left( \begin{array}{c} N-m-k-1 \\ l \end{array}
  \right)} \nn \\
&\times&
\sum_{j=0}^{N-m-k-l-1} { \left( \begin{array}{c} N-m-k-l-1 \\ j \end{array} \right)}
\sum_{i=0}^{N-m-k-l-j-1} { \left( \begin{array}{c} N-m-k-l-j-1 \\ i
    \end{array} \right)} \nn \\
&\times&
c_2^m c_3^k  c_4^l c_5^j c_6^i  c_1^{N-m-k-l-j-i-1} 
\sum_{n=1}^{6} n c_n \bar{b}_{N+m+2k+3l+4j+5i+n-2}(\infty)\,.
\label{subrun} 
\ea
The expressions for $S_{\eta}$ and the $c_i$ are now of course much
more complicated. Explicit expressions up to fifth order in the coupling
constant $\alpha_S(\mu^2)$ are presented in the Appendix. The fifth order
appears here as a consequence of using the two-loop expansion of the running
coupling and including the first four terms in the expansion of $S_\eta$ (\ref{eq:abseries}).

\begin{figure}[p]
\begin{center}
\mbox{\epsfig{figure=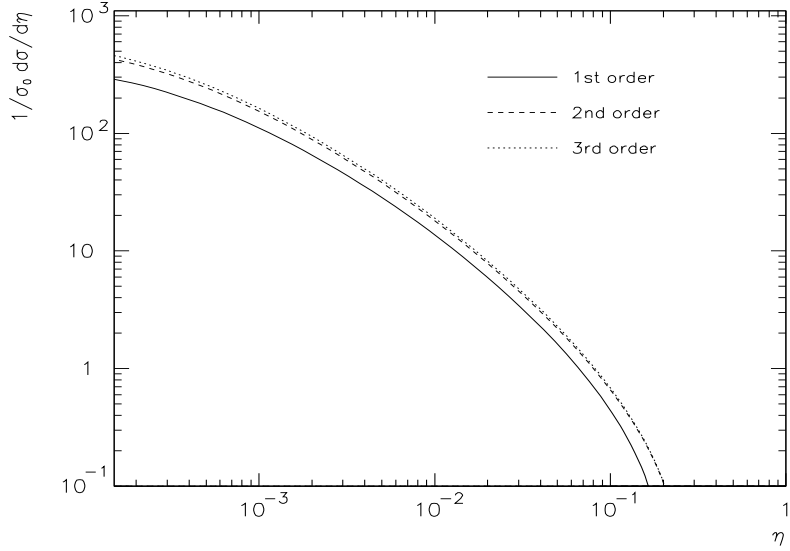,height=7.5cm,width=14cm}} 
\end{center}
\caption{Resummation of~(\ref{subrun}) for the first three orders in
  $\alpha_S(\mu^2)$ in the residual sum. Here $\mu^2=Q^{2/3 }q_T^{4/3}$, $Q=M_Z=91.187$ GeV, $\alpha_S(M_Z^2)=0.113.$}
\label{subrun123}
\end{figure}

\begin{figure}[p]
\begin{center}
\mbox{\epsfig{figure=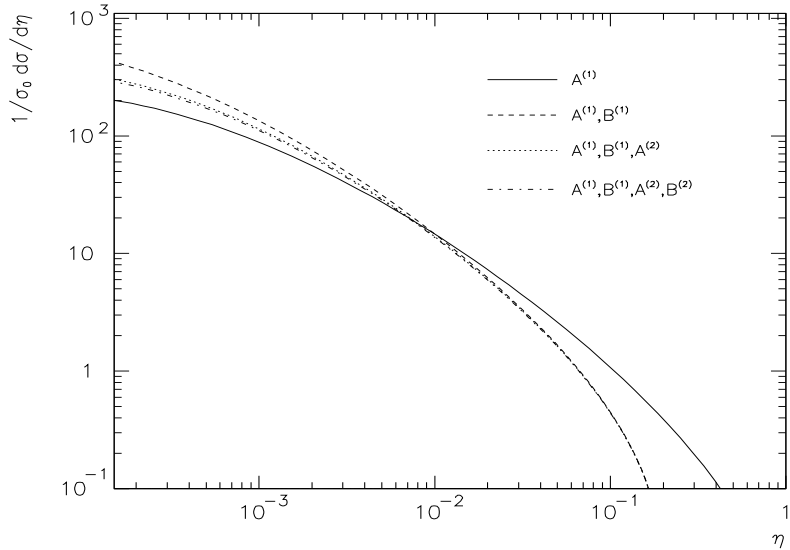,height=7.5cm,width=14cm}}
\end{center}
\caption{Resummation of ~(\ref{subrun}) for different subsets of nonzero
  coefficients, with
  only {$\cal O$}($\alpha_S(\mu^2)$) terms in the residual sum considered. 
  Here $\mu^2=Q^{2/3 }q_T^{4/3}$, $Q=M_Z=91.187$ GeV, $\alpha_S(M_Z^2)=0.113.$}
\label{subrun1}
\end{figure}

Numerical results based on the complete expression are displayed in 
Figs.~\ref{subrun123}-\ref{subrun1}, for the scale choice $\mu^2 = 
 Q^{2/3} q_T^{ 4 /3}$.  
First, 
Fig.~\ref{subrun123} shows the cross section with all four $A_i,\; B_i$ coefficients
included, together with the first three orders in $\alpha_S(\mu^2)$ in the
entire sum over $N$ in~(\ref{subrun}). 
Notice the rapid convergence when the higher-orders
are included.  Fig.~\ref{subrun1} shows how the first-order cross section
is influenced by the various $A_i,\; B_i$ coefficients. Notice how the relative 
impact on the leading-order  $A_1$ result is essentially $B_1 > A_2 > B_2$,
as might be expected. The effect of including $B_2$ is numerically small, 
indicating a reasonable degree of convergence from the higher-order coefficients.

\begin{figure}
\begin{center}
\mbox{\hspace{1.5cm}\epsfig{figure=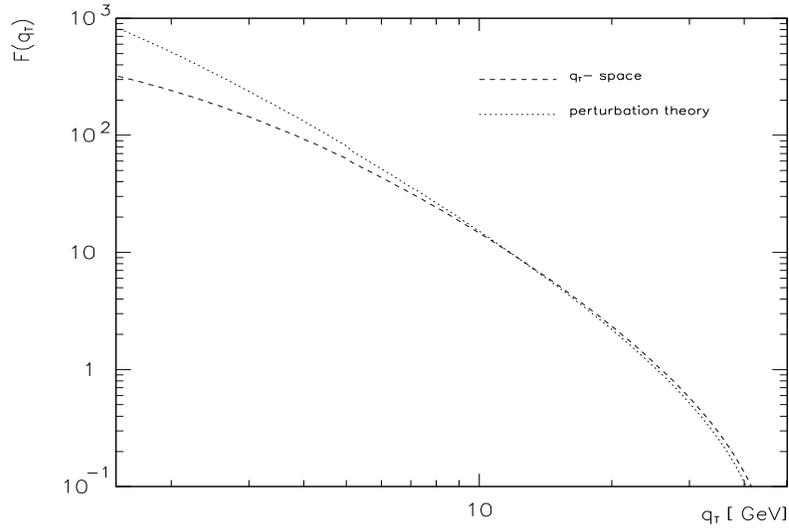,height=7.5cm,width=14cm}} 
\end{center}
\caption{Form factors $F^{(p)}(q_T)$, $F^{(q_T)}(q_T)$,
with $Q=M_Z=91.187$¬GeV, $\alpha_S(M_Z^2)=0.113$.} 
\label{Fpqt}
\end{figure}

The result~(\ref{subrun}) is valid in general for any choice of the
renormalization scale $\mu$. However, the expressions for $S_\eta$ and the $c$'s will
change depending on the particular scale choice. 
For $\mu^2=Q^2$ the expression for the
cross section  ${1\over \sigma_0} {d \sigma \over d \eta}$ corresponds to
$F^{(q_T)}(q_T)$ in Ref.~\cite{Ellis:Veseli}. 
More precisely,
$F^{(q_T)}(q_T)$ as defined in \cite{Ellis:Veseli}
 can be obtained from~(\ref{subrun}) by choosing an upper
limit of summation $N=5$ and putting $\bar{b}_i(\infty)=0$ for $i\geq 1$.
This recipe comes from the observation that  the only coefficient appearing
together with $\bar{b}_0(\infty)$ is $c_1$, which contains terms up to 
${\cal{O}}(\alpha_S^5(Q^2))$.%
\footnote{Note that throughout this work we consider the two-loop expansion 
  of the running coupling.}
Moreover, the expression~(\ref{subrun}), when
calculated for $\mu^2=Q^2$ and expanded in powers of $\alpha_S(Q^2)$ gives the 
${\cal{O}}(\alpha_S^2(Q^2))$ `perturbation theory' result 
 $F^{(p)}(q_T)$ from~\cite{Ellis:Veseli}. In Fig.~\ref{Fpqt} we show
$F^{(p)}(q_T)$ and $F^{(q_T)}(q_T)$ as functions of $q_T$, analogous to Fig.~3
in~\cite{Ellis:Veseli}.\footnote{Although we agree with~\cite{Ellis:Veseli} in the 
analytical result for the
perturbative expansion, there is a significant numerical discrepancy that we are
unable to account for.} 

\begin{figure}
\begin{center}
\mbox{\hspace{1.5cm}\epsfig{figure=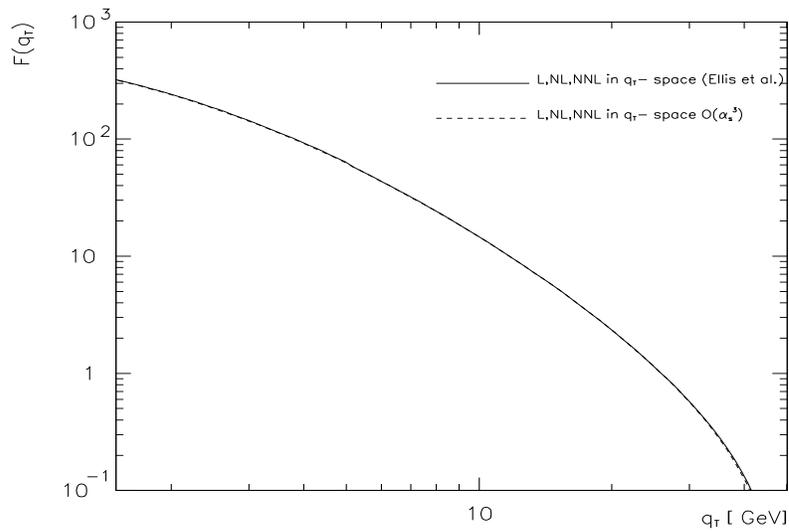,height=7.5cm,width=14cm}} 
\end{center}
\caption{Resummation of the L, NL and NNL `towers' of logarithms according
  to~(\ref{subrun}) for $\mu^2=Q^2$ versus
  $F^{(q_T)}(q_T)$ from~\cite{Ellis:Veseli}, with
  $Q=M_Z=91.187$ GeV, $\alpha_S(M_Z^2)=0.113$.} 
\label{NNLcomp}
\end{figure}

\begin{figure}
\begin{center}
\mbox{\epsfig{figure=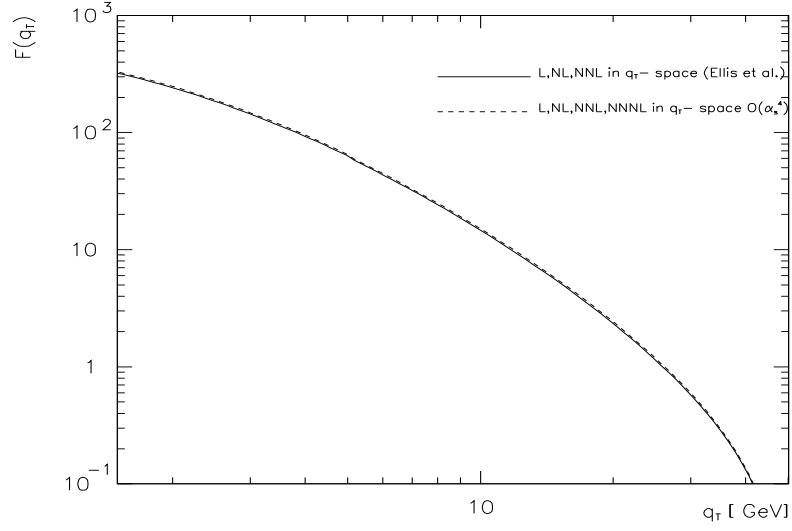,height=7.5cm,width=14cm}} 
\end{center}
\caption{Resummation of the first three (L,NL,NNL) `towers' according
  to~\cite{Ellis:Veseli} and the first four (L,NL,NNL,NNNL) towers
  of logarithms~(\ref{subrun}) for $\mu^2=Q^2$, $Q=M_Z=91.187$ GeV, $\alpha_S(M_Z^2)=0.113$.} 
\label{NNNL}
\end{figure}

\begin{figure}
\begin{center}
\mbox{\epsfig{figure=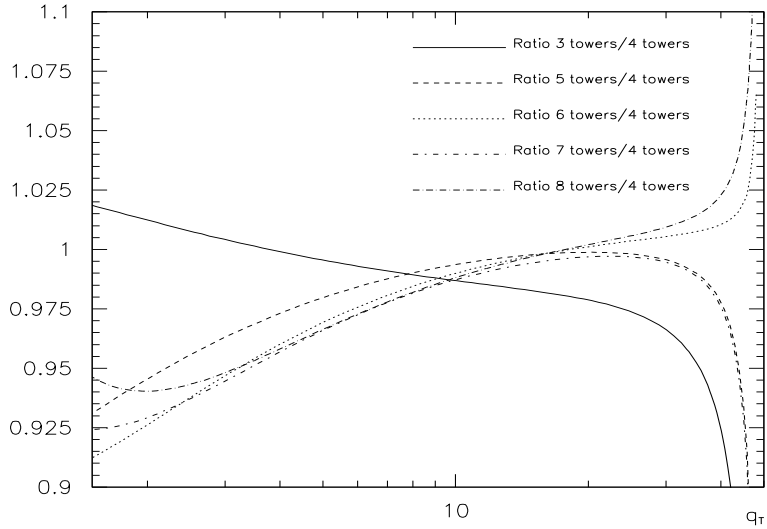,height=7.5cm,width=14cm}}
\end{center}
\caption{Ratio of the results for 3,5,6,7,8 `towers' of logarithms,
  normalised to the 4-th tower result, for $\mu^2=Q^{2/3}q_T^{4/3}$, $Q=M_Z=91.187$ GeV, $\alpha_S(M_Z^2)=0.113$.}   
\label{ratio}
\end{figure}

The $q_T$-space formalism as presented in~\cite{Ellis:Veseli} does not
take account of non-zero $\bar{b}_i(\infty)$ with $i\geq 1$,  and therefore
does not include known contributions from sub-leading 
NNNL `towers' of (kinematic) logarithms. This is partially compensated for
in~\cite{Ellis:Veseli} by
a redefinition  of $B^{(2)}$,
\be
\tilde{B}^{(2)}= B^{(2)} + 2 A^{(1)}\zeta(3)
\ee
which, although it  does  correctly account for a
${\cal{O}}(\alpha_S^2(Q^2))$ term from the NNNL `tower' (see Table~\ref{tow:ellis}),  
distorts other terms from this
 `tower'.\footnote{ We also find a  disagreement  with the curves in
  Fig.~4 of Ref.~\cite{Ellis:Veseli}. Since
  $\tilde{B}^{(2)} > B^{(2)} > 0$,
  after the replacement  of $ B^{(2)}$ by $\tilde{B}^{(2)}$ one would expect a
  {\it smaller} value for the cross section, in contrast to the displayed result
  in Ref.~\cite{Ellis:Veseli}.}
With the help of our expression~(\ref{subrun}) we can obtain the first {\it four}
`towers' fully resummed. It should however be remembered that the result is not 
`pure', in the sense that it  contains additional sub-leading terms.  
On the other hand, an effect induced by
the redefinition of $B^{(2)}$ seems to be comparable with that caused by
summing these sub-leading terms in~(\ref{subrun}) (with non-zero
  $\bar{b}_3(\infty)$, $\bar{b}_5(\infty) )$,
see Fig.~\ref{NNLcomp}. 
A difference obviously arises when the fourth tower
is also resummed (cf. Fig.~\ref{NNNL}) --- numerically we encounter
an increase in the cross section of approximately 3\% for all values of
$q_T$, when the scale equals $\mu^2=Q^2$. Furthermore, Fig.~\ref{ratio} shows the effect of including 3,5,6,7,8
`towers', normalised to the 4th-tower result, now for the scale choice
$\mu^2=Q^{2/3}q_T^{4/3}$. The figure clearly shows the numerical importance
of the kinematical logarithms of the higher towers, and also the stability
over a broad range of relevant $q_T$ as more towers are included.
However, this change is approximately of the same
magnitude as the one observed for the change of the renormalization scale. 
In particular, for the 4 towers of logarithms, changing the scale from our default $\mu^2=Q^{2/3}q_T^{4/3}$
to the (lower) scale $\mu^2=q_T^2$
and the (higher) scale $\mu^2=Q^2$
changes the $q_T$ distribution by less than $\pm 3\%$ over the complete
low-$q_T$ range.

\section{Summary and conclusions}
\label{sec:conc}

The $q_T$-space formalism for describing vector boson production in hadron
collisions is known to overcome many of the problems faced by the $b$-space method.
In this paper we have further investigated the $q_T$-space approach. For the
parton-level subprocess cross section we have modified the existing approach 
in order to incorporate sub-leading `kinematic' logarithms. 
We have carefully studied the effect of various sub-leading contributions: higher-order
perturbative coefficients, the running coupling, and the `kinematic' logarithms.
We have confirmed that the `kinematic' logarithms are particularly important at small $q_T$, 
where they serve to cancel the suppressing effect of the Sudakov form factor.   

Our technique enables us to resum the first four logarithmic `towers'
{\it including} the NNNL series, together with the first few sub-leading `kinematic'
logarithms. We have shown that that the most significant quantitative 
change in the predictions
for the cross section is caused by resumming the NNNL `tower'. 
The fact that the fourth `tower' only changes the cross section by about $3\%$
shows that the convergence of our expansion is certainly adequate for phenomenological
applications. We note that a drawback  of this method is an inability to select a
particular numbers of `towers' to be fully resummed.  
  
In this paper we have concentrated only on the perturbative   contributions
to the cross section. It is well known that in practice non-perturbative
effects (`$k_T$ smearing') are also important, and affect the $W$ $q_T$ distribution
in the very low $q_T$ region, see for example Ref.~\cite{Ellis:Ross:Veseli}.
These must be taken into account before assessing the impact of the sub-leading
logarithmic contributions on the physical cross section.
We will address these issues in a forthcoming study.
  
\noindent{\it Note added:} As this paper was nearing completion we became
aware of an interesting
 new study of soft gluon resummation, Ref.~\cite{Frixione:Nason:Ridolfi}, 
which addresses the same problem. In Ref.~\cite{Frixione:Nason:Ridolfi}, resummation in $b$-space
is studied, in particular the impact of factorially growing terms 
in the $\alpha_s$ expansion therein. A closed expression is subsequently
obtained for the corresponding $q_T$ space result,
which resums logarithms at the NLL level in the Sudakov exponent. The
kinematic logarithms are also treated differently, such that a singularity is 
encountered as $q_T \to q_T^{\rm crit}>0$. We will compare the phenomenological implications
of the two approaches in a forthcoming study.

\vspace{1cm}

\noindent{\bf Acknowledgments}\\
This work was supported in part by the EU Fourth Framework Programme
`Training and Mobility of Researchers', Network `Quantum Chromodynamics and
the Deep Structure of Elementary Particles', contract FMRX-CT98-0194 (DG 12 -
MIHT). A.K. gratefully acknowledges financial support received from the 
Overseas Research Students Award Scheme and the University of Durham. 

\newpage
 
%%%%%%%%%%%%%%%%%%%%%     REFERENCES     %%%%%%%%%%%%%%%%%%%%%%%%%%%%%%%%

\section*{Appendix}

In this appendix we list the expressions for $S_{\eta}$ and the $c_i$
coefficients in~(\ref{subrun}), for the choice of the renormalization scale
$\mu^2=Q^2\eta^{2/3}$. 

\bann
&&S_{\eta } = - { \frac {1}{2}} \lambda 
 \Bigg[   { \frac {1}{270}} \,{ \frac {\alpha_S
 (\mu ^{2})^{5}\,{\beta _{0}}^{4}\,A^{(2)}\,{\rm ln}\bigg( { 
\frac {1}{\eta }} \bigg)^{6}\,}{\pi ^{5}\,A^{(1)}}}  
 + \bigg ( - { \frac {1}{135}} \,
{ \frac {{\beta _{0}}^{3}\,\alpha_S (\mu ^{2})^{4}\,
A^{(2)}}{\pi ^{4}\,A^{(1)}}}  \\ 
& & - 4\,\pi \,\bigg({ 
\frac {1}{540}} \,{ \frac {{\beta _{0}}^{2}\,{\beta 
_{1}}\,A^{(2)}}{\pi ^{6}\,A^{(1)}}}  - { \frac {11
}{1620}} \,{ \frac {{ \beta _{0}}^{4}\,B^{(2)}}{\pi 
^{6}\,A^{(1)}}} \bigg)\,\alpha_S (\mu ^{2})^{5}
\bigg)\,{\rm ln}\bigg(
{ \frac {1}{\eta }} \bigg)^{5}  \\ 
& & + \bigg( { \frac {1}{12}} \,{ \frac {{\beta
 _{0}}^{2}\,\alpha_S (\mu ^{2})^{3}\,A^{(2)}}{\pi ^{3}\,A^{(1)}}
}  - 4\,\pi \,\bigg( - { \frac {1}{144}} \,
{ \frac {{\beta _{1}}^{2}\,A^{(2)}}{\pi ^{6}\,A^{(1)}}}  
+ { \frac {5}{216}} \,{ \frac {{
\beta _{0}}^{2}\,{\beta _{1}}\,B^{(2)}}{\pi ^{6}\,A^{(1)}}} \bigg)\,
\alpha_S (\mu ^{2})^{5} \\ 
 & & \mbox{} - 4\,\pi \,\bigg( - { \frac {1}{72}} \,
{ \frac {{\beta _{0}}\,{\beta _{1}}\,A^{(2)}}{\pi 
^{5}\,A^{(1)}}}  + { \frac {5}{216}} \,
{ \frac {{\beta _{0}}^{3}\,B^{(2)}}{\pi ^{5}\,A^{(1)}}} \bigg)\,
\alpha_S (\mu ^{2})^{4} + { \frac {1}{18}} \,
{ \frac {{\beta _{0}}^{2}\,\alpha_S (\mu ^{2})^{2}}{
\pi ^{2}}} \bigg){\rm ln}\bigg({ \frac {1}{\eta }} \bigg)^{4} \\ 
 & & \mbox{} + \bigg({ \frac {1}{9}} \,{ 
\frac {{\beta _{1}}^{2}\,\alpha_S (\mu ^{2})^{5}\,B^{(2)}}{\pi ^{5
}\,A^{(1)}}}  + { \frac {2}{9}} \,{ 
\frac {{\beta _{0}}\,{\beta _{1}}\,\alpha_S (\mu ^{2})^{4}\,B^{(2)}}{\pi ^{4}\,
A^{(1)}}}  + { \frac {1}{3}} \,
{ \frac {{\beta _{0}}^{2}\,\alpha_S (\mu ^{2})^{3}\,
B^{(2)}}{\pi ^{3}\,A^{(1)}}}  + { \frac {2}{9}} \,
{ \frac {\alpha_S (\mu ^{2})^{2}\,{\beta _{0}}^{2}\,
B^{(1 )}}{\pi ^{2}\,A^{(1)}}} \bigg)\\
& & \times \,{\rm ln}\bigg({ \frac {1}{\eta }} \bigg)^{3}
+ \bigg( - { \frac {1}{3}} \,{ \frac {{
\beta _{1}}\,\alpha_S (\mu ^{2})^{3}\,B^{(2)}}{\pi ^{3}\,A^{(1)}}
}  - 4\,\pi \,\bigg({ \frac {1}{12}} \,{ 
\frac {{\beta _{0}}\,B^{(2)}}{\pi ^{3}\,A^{(1)}}}  + 
{ \frac {1}{12}} \,{ \frac {{\beta _{1}
}\,B^{(1)}}{\pi ^{3}\,A^{(1)}}} \bigg)\,\alpha_S (\mu ^{2})^{2} \\ 
& & - 4\,
\pi \,\bigg( - { \frac {1}{8}} \,{ \frac {
A^{(2)}}{\pi ^{2}\,A^{(1)}}}  + { \frac {1}{12}} 
\,{ \frac {{\beta _{0}}\,B^{(1)}}{\pi ^{2}\,A^{(1)}}} \bigg)\,
\alpha_S (\mu ^{2}) 
\mbox{} + 1\bigg){\rm ln}\bigg({ \frac {1}{\eta }} \bigg)^{2}
\mbox{} + \bigg({ \frac {\alpha_S (\mu ^{2})\,B^{(2)}}{\pi
 \,A^{(1)}}}  + 2\,{ \frac {B^{(1)}}{A^{(1)}}} \bigg)
\,{\rm ln}\bigg({ \frac {1}{\eta }} \bigg) \bigg] \,,
\eann
%%%
\begin{eqnarray*}
\lefteqn{{c_{1}}= \left( {\vrule 
height1.36em width0em depth1.36em} \right. \!  \! { 
\frac {1}{18}} \,{ \frac {\beta_1^{2}\,B^{(2)}\,
{\rm ln}\bigg({ \frac {1}{\eta }} \bigg)^{2}}{\pi ^{5}\,
A^{(1)}}}  + { \frac {1}{30}} \,{ \frac {\bigg(
{ \frac {10}{9}} \,{ \frac {\beta_0^{2}
\,\beta_1\,B^{(2)}}{ A^{(1)}}}  + { \frac {5}{3}} 
\,{ \frac {\beta_1^{2}\,A^{(2)}}{A1^{(1)}}} \bigg)\,
{\rm ln}\bigg({ \frac {1}{\eta }} \bigg)^{3}}{\pi ^{5}}} } \\
 & & \mbox{} + { \frac {1}{30}} \,{ 
\frac {\bigg({ \frac {5}{27}} \,{ \frac {
\beta_0^{4}\,B^{(2)}}{A^{(1)}}}  + { \frac {10}{9}
} \,{ \frac {\beta_0^{2}\,\beta_1\,A^{(2)}}{A^{(1)}
}} \bigg)\,{\rm ln}\bigg({ \frac {1}{\eta }} \bigg)^{4}}{\pi ^{5}}
}  + { \frac {1}{162}} \,{ \frac {{\rm 
ln}({ \frac {1}{\eta }} )^{5}\,\beta_0^{4}\,A^{(2)}
}{\pi ^{5}\,A^{(1)}}}  \! \! \left. {\vrule 
height1.36em width0em depth1.36em} \right) \alpha_S (\mu ^{2})^{5}
\mbox{}   \\
 & & + \left(  \! { \frac {1}{9}} \,{ 
\frac {\beta_0\,\beta_1\,B^{(2)}\,{\rm ln}\bigg({ 
\frac {1}{\eta }} \bigg)^{2}}{\pi ^{4}\,A^{(1)}}}  + { 
\frac {1}{27}} \,{ \frac {\beta_0^{3}\,A^{(2)}\,
{\rm ln}\bigg({ \frac {1}{\eta }} \bigg)^{4}}{\pi ^{4}\,{\it 
A^{(1)}}}}  + { \frac {1}{30}} \,{ \frac {\bigg(
{ \frac {10}{9}} \,{ \frac {\pi \,\beta
 0^{3}\,B^{(2)}}{A^{(1)}}}  + { \frac {10}{3}} \,
{ \frac {\pi \,\beta_0\,\beta_1\,A^{(2)}}{A^{(1)}}
} \bigg)\,{\rm ln}\bigg({ \frac {1}{\eta }} \bigg)^{3}}{\pi ^{5}}} 
 \!  \right)  \alpha_S (\mu ^{2})^{4} \\
 & & \mbox{} +  \left(  \! { \frac {1}{3}} \,
{ \frac {\beta_1\,B^{(2)}\,{\rm ln}\bigg({ 
\frac {1}{\eta }} \bigg)}{\pi ^{3}\,A^{(1)}}}  + { 
\frac {1}{30}} \,{ \frac {\bigg(5\,{ \frac {
\pi ^{2}\,\beta_0^{2}\,B^{(2)}}{A^{(1)}}}  + 10\,
{ \frac {\pi ^{2}\,\beta_1\,A^{(2)}}{A^{(1)}}} \bigg)\,
{\rm ln}\bigg({ \frac {1}{\eta }} \bigg)^{2}}{\pi ^{5}}}  + 
{ \frac {1}{6}} \,{ \frac {\beta_0^{2}
\,A^{(2)}\,{\rm ln}\bigg({ \frac {1}{\eta }} \bigg)^{3}}{\pi 
^{3}\,A^{(1)}}}  \!  \right) \,\alpha_S (\mu ^{2})^{3}   \\
 & & +  \left( 
{\vrule height1.36em width0em depth1.36em} \right. \!  \! { \frac {1}{30}} \,{ \frac {\bigg(10\,
{ \frac {\pi ^{3}\,\beta_0\,B^{(2)}}{A^{(1)}}}  + 
10\,{ \frac {\pi ^{3}\,\beta_1\,B^{(1)}}{A^{(1)}}
} \bigg)\,{\rm ln}\bigg({ \frac {1}{\eta }} \bigg)}{\pi ^{5}}}  \\
 & & \mbox{} + { \frac {1}{30}} \,{ 
\frac {\bigg({ \frac {10}{3}} \,{ \frac {\pi
 ^{3}\,\beta_0^{2}\,B^{(1)}}{A^{(1)}}}  + 10\,{ 
\frac {\pi ^{3}\,\beta_0\,A^{(2)}}{A^{(1)}}}  + 10\,\pi ^{3}\,
\beta_1\bigg)\,{\rm ln}\bigg({ \frac {1}{\eta }} \bigg)^{2}}{\pi ^{
5}}}  + { \frac {1}{9}} \,{ \frac {
\beta_0^{2}\,{\rm ln}\bigg({ \frac {1}{\eta }} \bigg)^{3}}{\pi
 ^{2}}}  \! \! \left. {\vrule height1.36em width0em depth1.36em}
 \right) \alpha_S (\mu ^{2})^{2} \\
 & & \mbox{} +  \left(  \! { \frac {1}{30}} \,
{ \frac {\bigg(10\,{ \frac {\pi ^{4}\,\beta 
0\,B^{(1)}}{A^{(1)}}}  + 15\,{ \frac {\pi ^{4}\,
A^{(2)}}{A^{(1)}}} \bigg)\,{\rm ln}\bigg({ \frac {1}{\eta }
} \bigg)}{\pi ^{5}}}  + { \frac {1}{3}} \,{ 
\frac {\beta_0\,{\rm ln}\bigg({ \frac {1}{\eta }} \bigg)^{2}}{
\pi }}  + { \frac {1}{2}} \,{ \frac {
B^{(2)}}{\pi \,A^{(1)}}}  \!  \right) \,\alpha_S (\mu ^{2}) + 
{\rm ln}\bigg({ \frac {1}{\eta }} \bigg) + { 
\frac {B^{(1)}}{A^{(1)}}}\,, 
\end{eqnarray*}
%%%%
\begin{eqnarray*}
\lefteqn{{c_{2}}= \left( {\vrule 
height1.36em width0em depth1.36em} \right. \!  \! { 
\frac {1}{3}} \,{ \frac {\beta_1^{2}\,B^{(2)}\,
{\rm ln}\bigg({ \frac {1}{\eta }} \bigg)}{\pi ^{5}\,A^{(1)}}
}  + { \frac {13}{162}} \,{ \frac {
\beta_0^{4}\,A^{(2)}\,{\rm ln}\bigg({ \frac {1}{\eta }} 
\bigg)^{4}}{\pi ^{5}\,A^{(1)}}}  + { \frac {1}{30}} \,
{ \frac {\bigg(10\,{ \frac {\beta_0^{2}\,
\beta_1\,B^{(2)}}{A^{(1)}}}  + { \frac {35}{3}} \,
{ \frac {\beta_1^{2}\,A^{(2)}}{A^{(1)}}}\bigg )\,{\rm 
ln}\bigg({ \frac {1}{\eta }} \bigg)^{2}}{\pi ^{5}}} } \\
 & & \mbox{} + { \frac {1}{30}} \,{ 
\frac {\bigg({ \frac {20}{9}} \,{ \frac {
\beta_0^{4}\,B^{(2)}}{A^{(1)}}}  + { \frac {100}{9
}} \,{ \frac {\beta_0^{2}\,\beta_1\,A^{(2)}}{A^{(1)}}} \bigg)\,{\rm ln}\bigg({ \frac {1}{\eta }} \bigg)^{3}}{\pi ^{5
}}}  \! \! \left. {\vrule height1.36em width0em depth1.36em}
 \right) \alpha_S (\mu ^{2})^{5}\mbox{}   \\
 & & + \left(  \! { \frac {2}{3}} \,{ 
\frac {\beta_0\,\beta_1\,B^{(2)}\,{\rm ln}\bigg({ 
\frac {1}{\eta }} \bigg)}{\pi ^{4}\,A^{(1)}}}  + { 
\frac {1}{30}} \,{ \frac {\bigg(10\,{ 
\frac {\pi \,\beta_0^{3}\,B^{(2)}}{A^{(1)}}}  + { 
\frac {70}{3}} \,{ \frac {\pi \,\beta_0\,\beta_1\,
A^{(2)}}{A^{(1)}}}\bigg )\,{\rm ln}\bigg({ \frac {1}{\eta }
} \bigg)^{2}}{\pi ^{5}}}  + { \frac {10}{27}} \,
{ \frac {\beta_0^{3}\,A^{(2)}\,{\rm ln}\bigg(
{ \frac {1}{\eta }} \bigg)^{3}}{\pi ^{4}\,A^{(1)}}}  \! 
 \right)  \\
 & & \alpha_S (\mu ^{2})^{4}\mbox{} +  \left(  \! { 
\frac {\beta_1\,B^{(2)}}{\pi ^{3}\,A^{(1)}}}  + { \bigg(
\frac {1}{30}} \,{ \frac {\bigg(30\,{ 
\frac {\pi ^{2}\,\beta_0^{2}\,B^{(2)}}{A^{(1)}}}  + 40\,
{ \frac {\pi ^{2}\,\beta_1\,A^{(2)}}{A^{(1)}}} \bigg)\,
{\rm ln}\bigg({ \frac {1}{\eta }} \bigg)}{\pi ^{5}}}  + 
{ \frac {7}{6}} \,{ \frac {\beta_0^{2}
\,A^{(2)}\,{\rm ln}\bigg({ \frac {1}{\eta }} \bigg)^{2}}{\pi 
^{3}\,A^{(1)}}}  \!  \right) \,\alpha_S (\mu ^{2})^{3}  \\
 & &+  \left( 
{\vrule height1.36em width0em depth1.36em} \right. \!  \! 
 { \frac {1}{30}} \,{ \frac {30\,
{ \frac {\pi ^{3}\,\beta_0\,B^{(2)}}{A^{(1)}}}  + 
30\,{ \frac {\pi ^{3}\,\beta_1\,B^{(1)}}{A^{(1)}}
} }{\pi ^{5}}}  + { \frac {7}{9}} \,{ 
\frac {\beta_0^{2}\,{\rm ln}\bigg({ \frac {1}{\eta }} \bigg)^{
2}}{\pi ^{2}}}  \\
 & & \mbox{} + { \frac {1}{30}} \,{ 
\frac {\bigg(20\,{ \frac {\pi ^{3}\,\beta_0^{2}\,B^{(1)}
}{A^{(1)}}}  + 40\,{ \frac {\pi ^{3}\,\beta_0\,
A^{(2)}}{A^{(1)}}}  + 40\,\pi ^{3}\,\beta_1\bigg)\,{\rm ln}\bigg(
{ \frac {1}{\eta }} \bigg)}{\pi ^{5}}}  \! \! \left. 
{\vrule height1.36em width0em depth1.36em} \right) \alpha_S (\mu ^{
2})^{2} \\
 & & \mbox{} +  \left(  \! { \frac {1}{30}} \,
{ \frac {30\,{ \frac {\pi ^{4}\,\beta_0
\,B^{(1)}}{A^{(1)}}}  + 15\,{ \frac {\pi ^{4}\,
A^{(2)}}{A^{(1)}}} }{\pi ^{5}}}  + { \frac {4}{3}
} \,{ \frac {\beta_0\,{\rm ln}\bigg({ 
\frac {1}{\eta }} \bigg)}{\pi }}  \!  \right) \,\alpha_S (\mu ^{2}) + 1\,,
\end{eqnarray*}

\begin{eqnarray*}
\lefteqn{{c_{3}}= \left( {\vrule 
height1.36em width0em depth1.36em} \right. \!  \! { 
\frac {1}{30}} \,{ \frac {\bigg({ \frac {40
}{3}} \,{ \frac {\beta_0^{4}\,B^{(2)}}{A^{(1)}}} 
 + { \frac {160}{3}} \,{ \frac {\beta_0
^{2}\,\beta_1\,A^{(2)}}{A^{(1)}}} \bigg)\,{\rm ln}\bigg({ 
\frac {1}{\eta }}\bigg )^{2}}{\pi ^{5}}}  + { \frac {44}{
81}} \,{ \frac {\beta_0^{4}\,A^{(2)}\,{\rm ln}\bigg(
{ \frac {1}{\eta }} \bigg)^{3}}{\pi ^{5}\,A^{(1)}}} } \\
 & & \mbox{} + { \frac {1}{30}} \,{ 
\frac {\bigg(40\,{ \frac {\beta_0^{2}\,\beta_1\,B^{(2)}
}{A^{(1)}}}  + { \frac {100}{3}} \,{ 
\frac {\beta_1^{2}\,A^{(2)}}{A^{(1)}}} \bigg)\,{\rm ln}\bigg(
{ \frac {1}{\eta }} \bigg)}{\pi ^{5}}}  + { 
\frac {2}{3}} \,{ \frac {\beta_1^{2}\,B^{(2)}}{\pi 
^{5}\,A^{(1)}}}  \! \! \left. {\vrule 
height1.36em width0em depth1.36em} \right) \alpha_S (\mu ^{2})^{5}
 \\
 & & \mbox{} +  \left(  \! { \frac {4}{3}} \,
{ \frac {\beta_0\,\beta_1\,B^{(2)}}{\pi ^{4}\,{\it 
A^{(1)}}}}  + { \frac {16}{9}} \,{ \frac {
\beta_0^{3}\,A^{(2)}\,{\rm ln}\bigg({ \frac {1}{\eta }} 
\bigg)^{2}}{\pi ^{4}\,A^{(1)}}}  + { \frac {1}{30}} \,
{ \frac {\bigg(40\,{ \frac {\pi \,\beta_0^{3
}\,B^{(2)}}{A^{(1)}}}  + { \frac {200}{3}} \,
{ \frac {\pi \,\beta_0\,\beta_1\,A^{(2)}}{A^{(1)}}
} \bigg)\,{\rm ln}\bigg({ \frac {1}{\eta }} \bigg)}{\pi ^{5}}}  \! 
 \right) \,\alpha_S (\mu ^{2})^{4} \\
 & & \mbox{} +  \left(  \! { \frac {1}{30}} \,
{ \frac {60\,{ \frac {\pi ^{2}\,\beta_0
^{2}\,B^{(2)}}{A^{(1)}}}  + 40\,{ \frac {\pi ^{2}
\,\beta_1\,A^{(2)}}{A^{(1)}}} }{\pi ^{5}}}  + { 
\frac {10}{3}} \,{ \frac {\beta_0^{2}\,A^{(2)}\,
{\rm ln}\bigg({ \frac {1}{\eta }} \bigg)}{\pi ^{3}\,A^{(1)}}
}  \!  \right) \,\alpha_S (\mu ^{2})^{3} \\
 & & \mbox{} +  \left(  \! { \frac {1}{30}} \,
{ \frac {40\,{ \frac {\pi ^{3}\,\beta_0
^{2}\,B^{(1)}}{A^{(1)}}}  + 40\,{ \frac {\pi ^{3}
\,\beta_0\,A^{(2)}}{A^{(1)}}}  + 40\,\pi ^{3}\,\beta_1}{\pi ^{5
}}}  + { \frac {20}{9}} \,{ \frac {
\beta_0^{2}\,{\rm ln}\bigg({ \frac {1}{\eta }} \bigg)}{\pi ^{2
}}}  \!  \right) \,\alpha_S (\mu ^{2})^{2} + { \frac {
4}{3}} \,{ \frac {\beta_0\,\alpha_S (\mu ^{2})}{\pi }
}\,, 
\end{eqnarray*}
%%%%
\begin{eqnarray*}
\lefteqn{{c_{4}}= \left( {\vrule 
height1.36em width0em depth1.36em} \right. \!  \! { 
\frac {1}{30}} \,{ \frac {60\,{ \frac {
\beta_0^{2}\,\beta_1\,B^{(2)}}{A^{(1)}}}  + 30\,{ 
\frac {\beta_1^{2}\,A^{(2)}}{A^{(1)}}} }{\pi ^{5}}}  + 2\,
{ \frac {\beta_0^{4}\,A^{(2)}\,{\rm ln}\bigg(
{ \frac {1}{\eta }} \bigg)^{2}}{\pi ^{5}\,A^{(1)}}} } \\
 & & \mbox{} + { \frac {1}{30}} \,{ 
\frac {(40\,{ \frac {\beta_0^{4}\,B^{(2)}}{A^{(1)}
}}  + 120\,{ \frac {\beta_0^{2}\,\beta_1\,A^{(2)}}{
A^{(1)}}} )\,{\rm ln}\bigg({ \frac {1}{\eta }} \bigg)}{\pi ^{
5}}}  \! \! \left. {\vrule height1.36em width0em depth1.36em}
 \right) \alpha_S (\mu ^{2})^{5} \\
 & & \mbox{} +  \left(  \! { \frac {1}{30}} \,
{ \frac {60\,{ \frac {\pi \,\beta_0^{3}
\,B^{(2)}}{A^{(1)}}}  + 60\,{ \frac {\pi \,\beta_0
\,\beta_1\,A^{(2)}}{A^{(1)}}} }{\pi ^{5}}}  + 4\,
{ \frac {\beta_0^{3}\,A^{(2)}\,{\rm ln}\bigg(
{ \frac {1}{\eta }} \bigg)}{\pi ^{4}\,A^{(1)}}}  \! 
 \right) \,\alpha_S (\mu ^{2})^{4} + 3\,{ \frac {
\alpha_S (\mu ^{2})^{3}\,\beta_0^{2}\,A^{(2)}}{\pi ^{3}\,A^{(1)}}
}  \\
 & & \mbox{} + 2\,{ \frac {\beta_0^{2}\,\alpha_S (\mu 
^{2})^{2}}{\pi ^{2}}}\,, 
\end{eqnarray*}

\[
{c_{5}}= \left(  \! { \frac {56}{15}} \,
{ \frac {\beta_0^{4}\,A^{(2)}\,{\rm ln}\bigg(
{ \frac {1}{\eta }}\bigg )}{\pi ^{5}\,A^{(1)}}}  + 
{ \frac {1}{30}} \,{ \frac {48\,
{ \frac {\beta_0^{4}\,B^{(2)}}{A^{(1)}}}  + 96\,
{ \frac {\beta_0^{2}\,\beta_1\,A^{(2)}}{A^{(1)}}} 
}{\pi ^{5}}}  \!  \right) \,\alpha_S (\mu ^{2})^{5} + 
{ \frac {16}{5}} \,{ \frac {\alpha_S (\mu
 ^{2})^{4}\,\beta_0^{3}\,A^{(2)}}{\pi ^{4}\,A^{(1)}}} \,,
\]

\[
{c_{6}}={ \frac {8}{3}} \,{ \frac {
\beta_0^{4}\,A^{(2)}\,\alpha_S (\mu ^{2})^{5}}{\pi ^{5}\,A^{(1)}}
} \,.
\]

\clearpage
%%%%%%%%%%%%%%%%%%%%%%     FIGURES    %%%%%%%%%%%%%%%%%%%%%%%%%%%%%%%%%%
%%%%%%%%%%%%%%%%%%%%%%%%%%%%%%%%%%%%%%%%%%%%%%%%%%%%%%%%%%%%%%%%%%%%%%%%%%%
%%%%%%%%%%%%%%%%%%%%%%%%%%%%%%%%%%%%%%%%%%%%%%%%%%%%%%%%%%%%%%%%%%%%%%%%%%%
%%%%%%%%%%%%%%%%%%%%%%%%%%%%%%%%%%%%%%%%%%%%%%%%%%%%%%%%%%%%%%%%%%%%%%%%%%%%
\begin{table}[p]
\begin{center}
\begin{tabular}{|c|c|c|c| }           \hline \hline
 m &  $\bar b_m (\infty)$    &  m  &  $\bar b_m (\infty)$ \\ \hline 
 0 &  1.0                    &  10 &  1122.9875510 \\  
 1 &  0                      &  11 &  $-$6141.3046770\\
 2 &  0                      &  12 &  36851.269530\\
 3 &  $-$.601028451          &  13 &  $-$239674.372200\\
 4 &  0                      &  14 &  1677209.4750 \\
 5 &  $-$1.555391633         &  15 &  $-$12580409.1300 \\
 6 &  3.612351995            &  16 &  100640859.60 \\
 7 &  $-$11.343929370        &  17 &  $-$855451267.600\\
 8 &  52.350738970           &  18 &  7.699062951e+09 \\
 9 &  $-$218.6078590         &  19 &  $-$7.314109389e+10 \\ \hline \hline
\end{tabular}
\end{center}
\caption{The first 20 coefficients $\bar b_m({\infty})$, calculated according
to~(\ref{b_form}).}
\label{b_coeff}
\end{table}
%%%%%%%%%%%%%%%%%%%%%%%%%%%%%%%%%%%%%%%%%%%%%%%%%%%%%%%%%%%%%%%%%%%
%%%%%%%%%%%%%%%%%%%%%%%%%%%%%%%%%%%%%%%%%%%%%%%%%%%%%%%%%%%%%%%%%%%%%%%%%%%%%%%
%%%%%%%%%%%%%%%%%%%%%%%%%%%%%%%%%%%%%%%%%%%%%%%%%%%%%%%%%%%%%%%%%%%%%%%%%%%%
%%%%%%%%%%%%%%%%%%%%%%%%%%%%%%%%%%%%%%%%%%%%%%%%%%%%%%%%%%%%%%%%%%%%%%%%%%%%%
%%%%%%%%%%%%%%%%%%%%%%%%%%%%%%
%%%%%%%%%%%%%%%%
%%%%%%%%%%%
%%%%%%%%%%%%%%%%%%%%%%%%%%%%%%%%%%%%%%%%%%%%%%%%%%%%%%%%%%%%%%%%%%%%%%%%%%%%%
\clearpage 
\begin{table}   
\begin{center} 
\vspace{-2.5cm}
\begin{tabular}{|c||c|c|c| }      \hline  
&&&\\
6 &   0   &   0    &   0 \\
&&&\\
5 &   0   &   0               & ${1 \over 8} {A^{(1)}}^3$\\   
&&&\\
4 &   0   &   0                         & ${5 \over 8} {A^{(1)}}^2 B^{(1)}$\\ 
&&& \\ 
3 &   0   &$-{1 \over 2}{A^{(1)}}^2$
&                                 ${A^{(1)}} {B^{(1)}}^2-A^{(1)} A^{(2)}$
\\ 
&&&\\ 
2 &   0  & $-{3 \over 2}A^{(1)} B^{(1)}$  & $-{3 \over 2}B^{(1)} A^{(2)} +10
{A^{(1)}}^3 \bar b_3(\infty)+{1 \over 2}{B^{(1)}}^3 -{3 \over 2}A^{(1)}
B^{(2)}$  \\  
&&&\\  
1 & $A^{(1)}$ & $A^{(2)}-{B^{(1)}}^2$ & $-2B^{(1)}B^{(2)} +20{A^{(1)}}^2
B^{(1)} \bar  b_3(\infty)$ \\    
&&&\\
0 & $B^{(1)}$ & $-4{A^{(1)}}^2 \bar b_3(\infty) + B^{(2)}$ & $4{A^{(1)}}^3
\bar b_5(\infty) - 8A^{(1)} A^{(2)}\bar b_3(\infty) + 8A^{(1)}
{B^{(1)}}^2\bar b_3(\infty)$ \\  
&&&  \\ \hline\hline   
& 1 & 2 & 3 \\  \hline     
\end{tabular}
\end{center}  
\caption{Coefficients of the logarithmic terms in~(\ref{sublead}), with the Sudakov
factor expanded, for the first three orders in $ {\alpha_S\over 2
\pi}$. The rows correspond to powers of ${\alpha_S \over 2 \pi}$, the columns to 
powers of $\ln\bigg( {1 \over \eta}\bigg)$.}        
\label{tow:ellis}  
\end{table}  
%%%%%%%%%%%%%%%%%%%%%%
\begin{table}    
\begin{center}
\vspace{-.5cm}
\begin{tabular}{|c||c|c|c| }      \hline 
&&& \\
3 &   0   &  0   &  0 \\
&&&\\
2 &   0   &  0   & $8{A^{(1)}}^3\bar b_3(\infty)$ \\   
&&&\\
1 & $A^{(1)}$  & $A^{(2)}$   & $16{A^{(1)}}^2 B^{(1)}\bar b_3(\infty)$ \\
&&& \\
0 & $B^{(1)}$  & $-4{A^{(1)}}^2\bar b_3(\infty) +B^{(2)}$  &
$4{A^{(1)}}^3\bar b_5(\infty)- 8A^{(1)}A^{(2)}\bar b_3(\infty)+ 8A^{(1)}
{B^{(1)}}^2\bar b_3(\infty)$ 
\\ 
&&&\\ \hline\hline  
  & 1 & 2 & 3 \\ \hline
\end{tabular}
\end{center}
\caption{Coefficients of logarithmic terms in ~(\ref{sublead}) for
 the first three orders in  ${\alpha_S \over 2 \pi}$.
 The rows correspond to powers of ${\alpha_S \over 2 \pi}$, the columns to 
powers of $\ln\bigg( {1 \over \eta}\bigg)$.
   Because $\bar b_1(\infty)=\bar
b_2(\infty)=\bar b_4(\infty)=0$, the  coefficients of  $\bigg({\alpha_S 
\over 2\pi}\bigg)^2 \ln\bigg( {1 \over \eta}\bigg)^2$ and $\bigg({\alpha_S
\over  2\pi}\bigg)^3 \ln\bigg( {1 \over \eta}\bigg)^3$ are zero, and 
for higher
orders the biggest power of a  $\ln \bigg( {1 \over \eta}\bigg)$ logarithm
is equal to the order in ${\alpha_S \over 2 \pi}$.}    
\label{tow:we}
\end{table}
\clearpage
%%%%%%%%%%%%%%%%%%%%%%%%%%%%%%%%%%%%%%%%%%%%%%%%%%%%%%%%%%%%%%%%%%%%%%%%%%
\pagebreak
%%%%%%%%%%%%%%%%%%%%%%
%%%%%%%%%%%%%%%%%%%%%%%%%%%%%%%%%%%%%%%%%%%%%%%%%%%%%%%%%%%%%%%%%%%%%%%%%%%%%%%%%%
%%%%%%%%%%%%%%%%%%%%%%%%%%%%%%%%%%%%%%%%%%%%%%%%%%%%%%%%%%%%%%%%%%%%%%%%%%%%%%%%%%%% 
%%%%%%%%%%%%%%%%%%%%%%%%%%%%%%%%%%%%%%%%%%%%%%%%%%%%%%%%%%%%%%%%%%%%%%%%%%%%%%%%
%%%%%%%%%%%%%%%%%%%%%%%%%%%%%%%%%%%%%%%%%%%%%%%%%%%%%%%%%%%%%%%%%%%%%%%%%%%%%%%
%%%%%%%%%%%%%%%%%%%%%%%%%%%%%%%%%%%%%%%%%%%%%%%%%%%%%%%%%%%%%%%%%%%%%%%%%%%%%%%
%%%%%%%%%%%%%%%%%%%%%%%%%%%%%%%%%%%%%%%%%%%%%%%%%%%%%%%%%%%%%%%%%%%%%%%%%%%%%%%

\end{document}